\documentclass{osa-article}
\usepackage[latin9]{inputenc}
\usepackage{units}
\usepackage{amsmath}
\usepackage{graphicx}

\makeatletter

\newcommand{\lyxmathsym}[1]{\ifmmode\begingroup\def\b@ld{bold}
  \text{\ifx\math@version\b@ld\bfseries\fi#1}\endgroup\else#1\fi}

\journal{oe}

\articletype{Research Article}

\ifdefined\showcaptionsetup
 \PassOptionsToPackage{caption=false}{subfig}
\fi
\usepackage{subfig}
\makeatother

\begin{document}

\title{Entanglement mediated by DC current induced nonreciprocal graphene
plasmonics}

\author{Jay A. Berres,\authormark{1,{*}} S. Ali Hassani Gangaraj,\authormark{2,3}
and George W. Hanson\authormark{1}}

\address{\authormark{1}Department of Electrical Engineering, University of Wisconsin-Milwaukee, Milwaukee, Wisconsin 53211, USA\\
\authormark{2}Department of Electrical and Computer Engineering, University of Wisconsin-Madison, Madison, Wisconsin 53706, USA\\
\authormark{3}Optical Physics Division, Corning Research and Development Corporation, Sullivan Park, Corning, NY 14831, USA}

\email{\authormark{*}jaberres@uwm.edu} 


\begin{abstract}
We investigate entanglement mediated by DC current induced nonreciprocal
graphene plasmon polaritons. Nonreciprocal systems are ideal for the
enhancement, control, and preservation of entanglement due to the
potential for unidirectional beam-like wave propagation, i.e., efficiently
transporting photons from one emitter to another. Using a quantum
master equation and three-dimensional Green's function analysis, we
investigate a system consisting of two two-level emitters dominantly
interacting via electric current induced nonreciprocal plasmonic modes
of a graphene waveguide. We use concurrence as a measure of entanglement.
We show that nonreciprocal graphene plasmon polaritons are a promising
candidate to generate and mediate concurrence, where it is shown that
there is good enhancement and control of entanglement over vacuum,
which is beneficial for the broad applications of entanglement as
a quantum resource. We believe our findings contribute to the development
of quantum devices, enabling efficient and tunable entanglement between
two-level systems, which is a central goal in quantum technologies.
\end{abstract}

\section{Introduction}

The key element of any quantum system, e.g., a two-level system, is
for the quantum emitters to interact with the environment, and exchange
photons with other neighbor emitters. Tailoring these interactions
fundamentally alters the performance of quantum devices for maintaining
entanglement and quantum superposition between two separated quantum
emitters. Recent work has investigated entanglement mediated by surface
plasmon polaritons (SPPs) by means of various media and plasmonic
waveguide structures, e.g., V-shaped waveguides cut in a flat metal
plane \cite{GonzalezTudela2011,MartinCano2011,Gangaraj2015} and photonic
topological insulators (PTIs) \cite{Wang2008,Buhmann2012,Silveirinha2015,GonzalezBallestero2015,Gangaraj2017}.
It was shown that entanglement mediation could be enhanced, controlled,
and even preserved in the presence of large structural defects (by
means of the PTIs) by SPPs in the various plasmonic environments.
Additionally, in a recent work \cite{Gangaraj2022}, a theoretical
analysis is provided on why nonreciprocal photon transduction enhances
inter-atomic excitation transport efficiency. Here, we investigate
entanglement mediated by DC current induced nonreciprocal graphene
plasmon polaritons. Since nonreciprocal systems have the potential
for unidirectional (one-way) beam-like wave propagation, i.e., the
ability to efficiently transport photons from one emitter to another,
they are ideal for the enhancement, control, and preservation of entanglement
\cite{Lodahl2017}. The most traditional way to achieve this is based
on the magneto-optical effect, which requires biasing plasma-like
materials, e.g., semiconductors, with a static magnetic field. However,
due to the need for a strong external magnetic bias, this approach
is impractical and typically results in a weak nonreciprocal response
at the desired THz and optical frequencies. An alternative way to
achieve a nonreciprocal response is to bias certain conducting materials
(metals, degenerately doped semiconductors, and graphene) with a direct
electric current, where a sufficiently strong nonreciprocal effect
can be achieved utilizing graphene, given its high electrical conductivity\cite{Neto2009,Shishir2009,Dorgan2010,Ozdemir2015,Ramamoorthy2016,Yamoah2017,Sabbaghi2018,Sabbaghi2022}.
This results in nonreciprocal graphene plasmon polaritons \cite{Sabbaghi2015,VanDuppen2016,Morgado2017,Wenger2018,Morgado2018,CorreasSerrano2019,HassaniGangaraj2022},
which may be a promising candidate to further enhance entanglement
mediation.

In this paper, using a quantum master equation and Green's function
analysis, we investigate the dynamics of two two-level systems coupled
to each other via electric current induced nonreciprocal plasmonic
modes of a graphene waveguiding platform. We use concurrence as a
measure of entanglement. The theoretical model for this is provided
in Sec. \ref{sec:Theoretical-model}. In Sec. \ref{sec:DC-current-induced}
we establish the nonreciprocal response of the DC current biased graphene,
and in Sec. \ref{sec:Results} we compare the resulting concurrences
to determine whether or not there is an improvement in entanglement
enhancement and control. This allows for the further refinement and
comparison of the environments that best mediate and maintain entanglement,
which is beneficial for the broad applications of entanglement as
a quantum resource, including quantum computing \cite{Nielsen2000,Wilde2013}
and quantum cryptography \cite{Acin2011}.

\section{Theoretical model\label{sec:Theoretical-model}}

\subsection{Two-Qubit Entanglement}

For a system of two qubits, assuming general (reciprocal or nonreciprocal)
media with an external coherent drive (i.e., a laser pump) applied
to each qubit, the master equation is ($\rho_{s}(t)=\rho_{s}$) \cite{Gruner1996,Dung1998,Dung2000,Buhmann2012,Angelatos2015,Gangaraj2015,Gangaraj2017},
\begin{align}
\frac{\partial\rho_{s}(t)}{\partial t} & =-\frac{i}{\hbar}\left[\mathrm{H_{s}+\mathrm{V^{AF}},\rho_{s}}\right]+\underset{i=1,2}{\sum}\frac{\Gamma_{ii}}{2}\left(2\sigma_{i}\rho_{s}\sigma_{i}^{\dagger}-\sigma_{i}^{\dagger}\sigma_{i}\rho_{s}-\rho_{s}\sigma_{i}^{\dagger}\sigma_{i}\right)\label{eq:master equation}\\
 & +\left(\frac{\Gamma_{21}}{2}+ig_{21}\right)\left(\sigma_{2}\rho_{s}\sigma_{1}^{\dagger}-\rho_{s}\sigma_{1}^{\dagger}\sigma_{2}\right)+\left(\frac{\Gamma_{21}}{2}-ig_{21}\right)\left(\sigma_{1}\rho_{s}\sigma_{2}^{\dagger}-\sigma_{2}^{\dagger}\sigma_{1}\rho_{s}\right)\nonumber \\
 & +\left(\frac{\Gamma_{12}}{2}+ig_{12}\right)\left(\sigma_{1}\rho_{s}\sigma_{2}^{\dagger}-\rho_{s}\sigma_{2}^{\dagger}\sigma_{1}\right)+\left(\frac{\Gamma_{12}}{2}-ig_{12}\right)\left(\sigma_{2}\rho_{s}\sigma_{1}^{\dagger}-\sigma_{1}^{\dagger}\sigma_{2}\rho_{s}\right),\nonumber 
\end{align}
where, 
\begin{align}
\mathrm{H}_{s} & =\underset{i=1,2}{\sum}\hbar\triangle\omega_{i}\sigma_{i}^{\dagger}\sigma_{i},
\end{align}
and
\begin{align}
\mathrm{V^{AF}} & =-\hbar\left(\mathrm{\varOmega}_{1}e^{-i\triangle_{l}t}\sigma_{1}^{\dagger}+\mathrm{\varOmega}_{1}^{\ast}e^{i\triangle_{l}t}\sigma_{1}\right)-\hbar\left(\mathrm{\varOmega}_{2}e^{-i\triangle_{l}t}\sigma_{2}^{\dagger}+\mathrm{\varOmega}_{2}^{\ast}e^{i\triangle_{l}t}\sigma_{2}\right).
\end{align}
$\mathrm{H}_{s}$ is the Hamiltonian of the decoupled qubits, where
$\triangle\omega_{i}=\omega_{0}-\omega_{l}-\delta_{i}$, with $\delta_{i}=g_{ii}$
being the Lamb shift. Since the Lamb shift for optical emitters is
typically on the order of a few GHz the effect of the Lamb shift for
optical frequencies is small ($\omega_{i}\thicksim10^{12}$ Hz, $\delta_{i}\thicksim10^{9}$
Hz), therefore the Lamb shift can be ignored, or assumed to be accounted
for in the definition of the transition angular frequency $\omega_{0}$,
i.e., $\triangle\omega_{i}=\omega_{0}-\omega_{l}-\delta_{i}$ becomes
$\triangle\omega_{i}\approx\omega_{0}-\omega_{l}$. $\mathrm{V^{AF}}$
is the external coherent drive (i.e., a laser pump) applied to each
qubit, where $\triangle_{l}=\omega_{0}-\omega_{l}$, which is the
detuning parameter, with $\omega_{l}$ being the laser angular frequency,
and $\mathrm{\varOmega}_{i}=\nicefrac{(\mathbf{d}\cdot\mathbf{E}_{0}^{i})}{\hbar}$,
which is a Rabi frequency. Here the external coherent drive field
is treated as a classical number given its large amplitude.

Assuming $\omega_{0}=\omega_{l}$ and $\mathrm{\varOmega}_{i}=\mathrm{\varOmega}_{i}^{\ast}$,
Eq. (\ref{eq:master equation}) then becomes
\begin{align}
\frac{\partial\rho_{s}(t)}{\partial t} & =i\mathrm{\varOmega}_{1}\left(\mathrm{\left(\sigma_{1}^{\dagger}\rho_{s}+\sigma_{1}\rho_{s}\right)}-\left(\rho_{s}\sigma_{1}^{\dagger}+\rho_{s}\sigma_{1}\right)\right)+i\mathrm{\mathrm{\varOmega}_{2}\left(\left(\sigma_{2}^{\dagger}\rho_{s}+\sigma_{2}\rho_{s}\right)-\left(\rho_{s}\sigma_{2}^{\dagger}+\rho_{s}\sigma_{2}\right)\right)}\label{eq:master equation reduced}\\
 & +\frac{\Gamma_{11}}{2}\left(2\sigma_{1}\rho_{s}\sigma_{1}^{\dagger}-\sigma_{1}^{\dagger}\sigma_{1}\rho_{s}-\rho_{s}\sigma_{1}^{\dagger}\sigma_{1}\right)+\frac{\Gamma_{22}}{2}\left(2\sigma_{2}\rho_{s}\sigma_{2}^{\dagger}-\sigma_{2}^{\dagger}\sigma_{2}\rho_{s}-\rho_{s}\sigma_{2}^{\dagger}\sigma_{2}\right)\nonumber \\
 & +\left(\frac{\Gamma_{21}}{2}+ig_{21}\right)\left(\sigma_{2}\rho_{s}\sigma_{1}^{\dagger}-\rho_{s}\sigma_{1}^{\dagger}\sigma_{2}\right)+\left(\frac{\Gamma_{21}}{2}-ig_{21}\right)\left(\sigma_{1}\rho_{s}\sigma_{2}^{\dagger}-\sigma_{2}^{\dagger}\sigma_{1}\rho_{s}\right)\nonumber \\
 & +\left(\frac{\Gamma_{12}}{2}+ig_{12}\right)\left(\sigma_{1}\rho_{s}\sigma_{2}^{\dagger}-\rho_{s}\sigma_{2}^{\dagger}\sigma_{1}\right)+\left(\frac{\Gamma_{12}}{2}-ig_{12}\right)\left(\sigma_{2}\rho_{s}\sigma_{1}^{\dagger}-\sigma_{1}^{\dagger}\sigma_{2}\rho_{s}\right),\nonumber 
\end{align}
where 
\begin{align}
\Gamma_{\alpha\beta}(\omega_{0}) & =\frac{2}{\varepsilon_{0}\hbar}\mathrm{Im}\left\{ \mathbf{d}\cdot\underline{\mathbf{G}}(\mathbf{r}_{\alpha},\mathbf{r}_{\beta},\omega_{0})\cdot\mathbf{d}\right\} ,\label{eq:decay rate}\\
g_{\alpha\beta}(\omega_{0}) & =\frac{1}{\varepsilon_{0}\hbar}\mathrm{Re}\left\{ \mathbf{d}\cdot\underline{\mathbf{G}}(\mathbf{r}_{\alpha},\mathbf{r}_{\beta},\omega_{0})\cdot\mathbf{d}\right\} .\label{eq:emitter transition frequency shift}
\end{align}
In Eqs. (\ref{eq:decay rate}) and (\ref{eq:emitter transition frequency shift}),
$\mathbf{d}$ is the atom (qubit) transition dipole moment, $\Gamma_{\alpha\alpha}$
and $\Gamma_{\alpha\beta}$ ($\alpha\neq\beta$) are the dissipative
decay rates of qubit $\alpha$ due to its interaction with the environment
and its interaction with qubit $\beta$ through the environment, $g_{\alpha\beta}$
($\alpha\neq\beta$) is the qubits' transition frequency shift induced
by dipole-dipole coupling, and $\underline{\mathbf{G}}(\mathbf{r}_{\alpha},\mathbf{r}_{\beta},\omega_{0})$
is the Green's tensor representing the environment. Note that $\Gamma_{12}=\Gamma_{21}$
and $g_{12}=g_{21}$ for the reciprocal case, and for identical emitters
(qubits), which we assume in this work, $\Gamma_{11}=\Gamma_{22}$.

\subsection{Green's function}

We solve for the Green's tensor that satisfies \cite{Gruner1996,Dung1998,Dung2000,Angelatos2015,Gangaraj2015},
\begin{align}
\nabla\times\nabla\times\underline{\mathbf{G}}(\mathbf{r},\mathbf{r}^{\prime},\omega) & -k_{0}^{2}\mu_{r}(\mathbf{r},\omega)\varepsilon_{r}(\mathbf{r},\omega)\underline{\mathbf{G}}(\mathbf{r},\mathbf{r}^{\prime},\omega)=k_{0}^{2}\underline{\mathbf{I}}\delta(\mathbf{r}-\mathbf{r}^{\prime}),\label{eq:Green tensor wave equation}
\end{align}
where $\mathbf{r},\mathbf{r}^{\prime}$ are the observation and source
point vectors, respectively, $k_{0}=\nicefrac{\omega}{c}=\omega\sqrt{\mu_{0}\varepsilon_{0}}$
is the vacuum wavenumber, $\omega$ is the angular frequency, $c$
is the speed of light in vacuum, $\mu_{r}(\mathbf{r},\omega)$ is
the relative permeability, $\varepsilon_{r}(\mathbf{r},\omega)$ is
the relative permittivity, and $\underline{\mathbf{I}}$ is the unit
3-by-3 tensor. We assume that we are working with non-magnetic materials
so we set $\mu_{r}(\mathbf{r},\omega)=1$. Additionally, throughout
this work we assume time-harmonic fields with time variations of the
form $e^{-i\omega t}$.

\begin{figure}[tp]
\begin{centering}
\includegraphics[width=8cm]{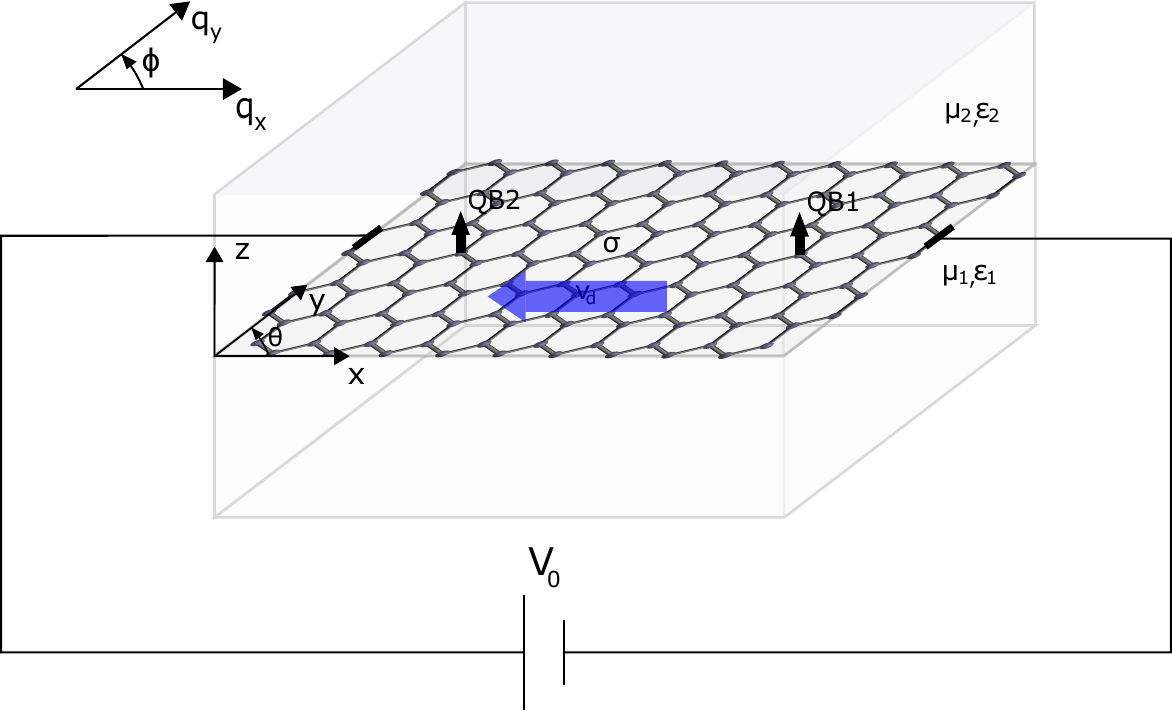} 
\par\end{centering}
\caption{Model for the case of a single interface (in this case an infinite
sheet of graphene modeled as an infinitesimally thin local two-sided
surface characterized by a surface conductivity $\sigma$) between
two different materials ($\mu_{2}$,$\varepsilon_{2}$ for $z>0$,
$\mu_{1}$,$\varepsilon_{1}$ for $z<0$). The graphene sheet is biased
with a DC voltage, where the blue arrow represents the drift velocity
induced by the DC current.\label{fig:Model-for-single-interface}}
\end{figure}

The model for the case of a single interface (in this case a sheet
of graphene modeled as an infinitesimally thin local two-sided surface
characterized by a surface conductivity $\sigma$) between two different
materials ($\mu_{2}$,$\varepsilon_{2}$ for $z>0$, $\mu_{1}$,$\varepsilon_{1}$
for $z<0$) \cite{Hanson2008} is seen in Fig. \ref{fig:Model-for-single-interface}.
We assume the interface is an infinite plane (in this case an infinite
sheet of graphene), where we also assume the source point and the
observation point are both above the interface, i.e., in region 2.
Additionally, we assume that the two atoms (qubits) have a dipole
moment of $\mathbf{d}=\widehat{\mathbf{z}}d$, i.e., they are polarized
perpendicular to the interface, with qubit one (QB1) at ($x^{\prime}$,$y^{\prime}$,$z^{\prime}$),
and qubit two (QB2) at ($x$,$y$,$z$), so that $\mathbf{R}=\mathbf{r}-\mathbf{r}^{\prime}=\left(x-x^{\prime}\right)\widehat{\mathbf{x}}+\left(y-y^{\prime}\right)\widehat{\mathbf{y}}+\left(z-z^{\prime}\right)\widehat{\mathbf{z}}$,
where $R=\left|\mathbf{R}\right|=\sqrt{\left(x-x^{\prime}\right)^{2}+\left(y-y^{\prime}\right)^{2}+\left(z-z^{\prime}\right)^{2}}$,
and with $\rho=\sqrt{\left(x-x^{\prime}\right)^{2}+\left(y-y^{\prime}\right)^{2}}$,
then, $R=\left|\mathbf{R}\right|=\sqrt{\rho^{2}+\left(z-z^{\prime}\right)^{2}}$.
We also assume that the qubits are at the same height, i.e., $z=z^{\prime}$,
which leads to $R=\rho$, and that QB1 is always located at the origin
($x^{\prime}=0$,$y^{\prime}=0$), i.e., at the center of the graphene
sheet ($x$-$y$ plane). We can apply a DC voltage to the graphene
sheet as shown in Fig. \ref{fig:Model-for-single-interface}, where
a drift velocity will be induced by the DC current.

For this geometry we can express the solution to Eq. (\ref{eq:Green tensor wave equation})
as
\begin{equation}
\underline{\mathbf{G}}(\mathbf{r},\mathbf{r}^{\prime},\omega)=\left[\underline{\mathbf{I}}k_{2}^{2}+\mathbf{\nabla}\mathbf{\nabla}\cdot\right]\left\{ \underline{\mathbf{g}}^{p}(\mathbf{r},\mathbf{r}^{\prime},\omega)+\underline{\mathbf{g}}^{s}(\mathbf{r},\mathbf{r}^{\prime},\omega)\right\} ,\label{eq:Green's function solution}
\end{equation}
where we are solving for the Green's function in region 2, therefore
we use $k_{2}=\omega\sqrt{\mu_{2}\varepsilon_{2}}$ as the wavenumber.
Here, $\underline{\mathbf{g}}^{p}(\mathbf{r},\mathbf{r}^{\prime},\omega)$
is the principle Green's function (the solution to Eq. (\ref{eq:Green tensor wave equation})
when $\varepsilon_{r}(\mathbf{r},\omega)=\varepsilon_{r2}(\mathbf{r},\omega)$,
i.e., the Green's function for a homogeneous medium (a single region
(region 2) with no interface)), and $\underline{\mathbf{g}}^{s}(\mathbf{r},\mathbf{r}^{\prime},\omega)$
is the scattered Green's function (the solution to Eq. (\ref{eq:Green tensor wave equation})
accounting for the field scattered from the media (the interface and
the regions)). With $\mathbf{d}=\widehat{\mathbf{z}}d$, then $\mathbf{d}\cdot\underline{\mathbf{G}}(\mathbf{r},\mathbf{r}^{\prime})\cdot\mathbf{d}$,
which is ultimately what we will need in Eqs. (\ref{eq:decay rate})
and (\ref{eq:emitter transition frequency shift}), becomes (using
the expression for $\underline{\mathbf{G}}(\mathbf{r},\mathbf{r}^{\prime},\omega)$
from Eq. (\ref{eq:Green's function solution}))
\begin{equation}
\mathbf{d}\cdot\underline{\mathbf{G}}(\mathbf{r},\mathbf{r}^{\prime})\cdot\mathbf{d}=d^{2}G_{zz}=d^{2}\left[k_{2}^{2}\left(g_{zz}^{p}+g_{zz}^{s}\right)+\frac{\partial^{2}}{\partial z^{2}}\left(g_{zz}^{p}+g_{zz}^{s}\right)\right],
\end{equation}
where
\begin{equation}
g_{zz}^{p}=\frac{e^{ik_{2}R}}{4\pi R}=\frac{e^{ik_{2}\sqrt{\rho^{2}+\left(z-z^{\prime}\right)^{2}}}}{4\pi\sqrt{\rho^{2}+\left(z-z^{\prime}\right)^{2}}},
\end{equation}
and \cite{Hanson2008}
\begin{align}
g_{zz}^{s} & =\frac{1}{\left(2\pi\right)^{2}}\int_{0}^{\infty}\int_{-\pi}^{\pi}R_{n}e^{-p_{2}\left(z+z^{\prime}\right)}\frac{e^{iq\rho\cos\left(\phi-\theta\right)}}{2p_{2}}qd\phi dq,\label{eq:scattered Green's function, double integral}
\end{align}
where,
\begin{equation}
R_{n}=\frac{\frac{\varepsilon_{1}}{\varepsilon_{2}}p_{2}-p_{1}+\frac{\sigma_{d}p_{2}p_{1}}{-i\omega\varepsilon_{2}}}{\frac{\varepsilon_{1}}{\varepsilon_{2}}p_{2}+p_{1}+\frac{\sigma_{d}p_{2}p_{1}}{-i\omega\varepsilon_{2}}}=\frac{N^{E}}{Z^{E}},\label{eq:scattering coefficient}
\end{equation}
$\varepsilon_{1}=\varepsilon_{0}\varepsilon_{r1}$, $\varepsilon_{2}=\varepsilon_{0}\varepsilon_{r2}$,
$p_{1}=\sqrt{q^{2}-k_{1}^{2}}$, and $p_{2}=\sqrt{q^{2}-k_{2}^{2}}$.
Note that we can also express $e^{iq\rho\cos\left(\phi-\theta\right)}$
in Eq. (\ref{eq:scattered Green's function, double integral}) as
$e^{iq\left(\cos(\phi)x+\sin(\phi)y\right)}$, where in both cases
$\phi$ is defined as the angle in the momentum coordinate space shown
in Fig. \ref{fig:Model-for-single-interface}. We will define a nonlocal
surface conductivity $\sigma_{d}$ in Sec. \ref{subsec:Nonlocal-Surface-Conductivity},
where we will see that for a drift velocity $v_{d}=0$, $\sigma_{d}=\sigma$,
defined in Sec. \ref{subsec:Local-surface-conductivity}. In that
case, $\sigma_{d}\rightarrow\sigma$, we can use a Bessel function
identity for the $\phi$ integral for $g_{zz}^{s}$, leading to
\begin{equation}
g_{zz}^{s}=\frac{1}{2\pi}\int_{0}^{\infty}R_{n}(\sigma_{d}\rightarrow\sigma)e^{-p_{2}\left(z+z^{\prime}\right)}\frac{J_{0}(q\rho)}{2p_{2}}qdq.
\end{equation}

To obtain the electric field from the Green's function we use
\begin{align}
E_{z} & =\frac{\mu_{r2}G_{zz}}{-i\omega\varepsilon_{0}}=\frac{G_{zz}}{-i\omega\varepsilon_{0}}.\label{eq:electric field}
\end{align}

\subsection{Local surface conductivity for graphene\label{subsec:Local-surface-conductivity}}

We define the local surface conductivity $\sigma$ for graphene as
(in the low-temperature limit) \cite{Hanson2015}
\begin{equation}
\sigma(\omega)=\frac{ie^{2}\mu_{c}}{\pi\hbar^{2}\left(\omega+i\Gamma\right)}+\frac{e^{2}}{4\hbar}\left[\Theta\left(\hbar\omega-2\mu_{c}\right)+\frac{i}{\pi}\ln\left|\frac{\hbar\omega-2\mu_{c}}{\hbar\omega+2\mu_{c}}\right|\right],\label{eq:local conductivity}
\end{equation}
where $\Theta\left(x\right)$ is the Heaviside function and $e$ is
the charge of an electron.

\begin{figure}[tp]
\begin{centering}
\hfill{}\subfloat[\label{fig:local surface conductivity subfigure}]{\centering{}\includegraphics[width=6cm,keepaspectratio,height=5cm]{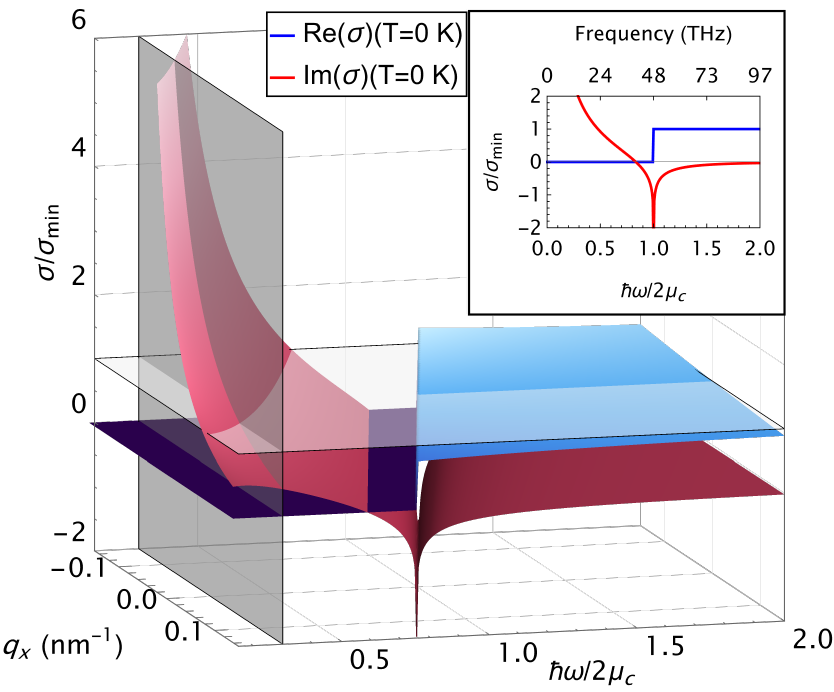}}\hfill{}\subfloat[\label{fig:nonlocal conductivity for 15 THz subfigure}]{\begin{centering}
\begin{minipage}[b][5cm][c]{6cm}%
\begin{center}
\includegraphics[width=6cm,keepaspectratio,height=5cm]{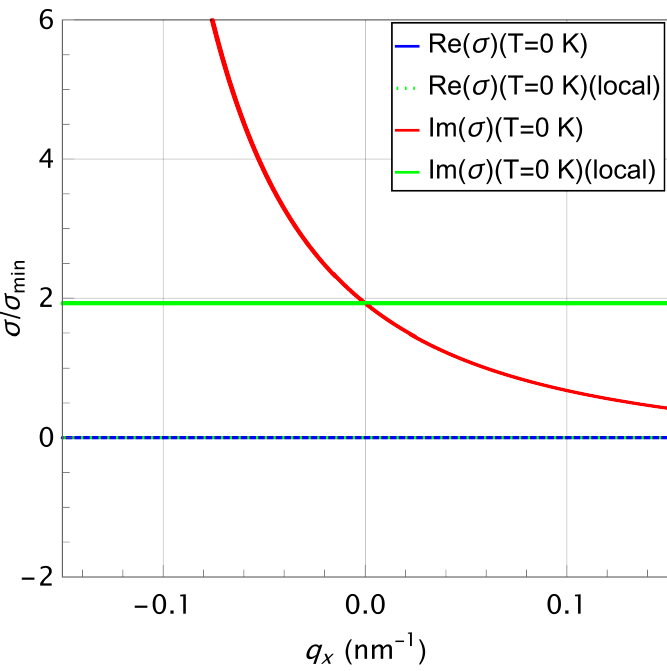}
\par\end{center}%
\end{minipage}
\par\end{centering}
}\hfill{}
\par\end{centering}
\caption{Nonlocal surface conductivity $\sigma_{d}(v_{d},q_{x},\omega)$ for
graphene (in the low-temperature limit), where $\mu_{c}=0.1$ eV,
$\tau=0.35$ ps, and $\sigma_{\mathrm{min}}=\nicefrac{\left(\pi e^{2}\right)}{\left(2h\right)}$
($h$ is Planck's constant): (a) nonlocal conductivity for $v_{d}=\nicefrac{-v_{F}}{2}$,
with inset of local surface conductivity $\sigma(\omega)$, and (b)
nonlocal conductivity for $v_{d}=\nicefrac{-v_{F}}{2}$ at 15 THz
(on the plane at 15 THz in (a)), where the local conductivity at 15
THz is also included for reference. A plane is included in (a) at
$\nicefrac{\sigma}{\sigma_{\mathrm{min}}}=1$ for reference to better
see changes to $\nicefrac{\sigma}{\sigma_{\mathrm{min}}}$ as $q_{x}$
changes.\label{fig:Local-nonlocal-surface-conductivity}}
\end{figure}

In Eq. (\ref{eq:local conductivity}), $\mu_{c}$ is the chemical
potential (Fermi energy) and $\Gamma$ is the phenomenological intraband
scattering rate, where $\Gamma=\nicefrac{1}{\tau}$ ($\tau$ is the
intraband scattering time). In this work we use \cite{Morgado2018}
$\mu_{c}=0.1$ eV and $\tau=0.35$ ps. Then, using these values, we
obtain the plot inset in Fig. \ref{fig:local surface conductivity subfigure}
for the local surface conductivity $\sigma(\omega)$ for graphene.
Additionally, since we are only interested in TM surface modes we
need to ensure that we are working at a frequency where the imaginary
part of $\sigma(\omega)$ (or the conductivity with the Doppler-shifted
frequency used in the $\sigma_{d}$ expression in Sec. \ref{subsec:Nonlocal-Surface-Conductivity})
is positive \cite{Hanson2015}.

\subsection{Concurrence}

We use the expression for concurrence in the general case (reciprocal
or nonreciprocal) \cite{Wootters2001,Gangaraj2017},
\begin{align}
C(t) & =\mathrm{max}\left(0,\sqrt{u_{1}}-\sqrt{u_{2}}-\sqrt{u_{3}}-\sqrt{u_{4}}\right),
\end{align}
as a measure of the amount of entanglement between the qubits (varies
from 0 (unentangled) to 1 (completely entangled), where $u_{i}$ are
the eigenvalues, arranged in descending order, of the matrix $\rho(t)\rho^{y}(t)$,
where $\rho^{y}(t)=\sigma_{y_{1}}\otimes\sigma_{y_{2}}\rho^{\ast}(t)\sigma_{y_{1}}\otimes\sigma_{y_{2}}$
is the spin-flip density matrix with $\sigma_{y_{i}}$ being the Pauli
matrix, $\sigma_{y_{i}}=\begin{bmatrix}\begin{array}{cc}
0 & -i\\
i & 0
\end{array}\end{bmatrix}$, in the spin-$1/2$ basis. Expressing $\sigma_{y_{i}}$ in terms
of the operator basis we are using, the atomic transition operators,
$\sigma_{i}^{+}=\sigma_{i}^{\dagger}=\bigl|e_{i}\bigr\rangle\bigl\langle g_{i}\bigr|,\:\sigma_{i}^{-}=\sigma_{i}=\bigl|g_{i}\bigr\rangle\bigl\langle e_{i}\bigr|$,
where we have defined $\bigl|g_{i}\bigr\rangle=\begin{bmatrix}\begin{array}{c}
1\\
0
\end{array}\end{bmatrix},\bigl|e_{i}\bigr\rangle=\begin{bmatrix}\begin{array}{c}
0\\
1
\end{array}\end{bmatrix}$, we have $\sigma_{y_{i}}=i\left(\sigma_{i}^{\dagger}-\sigma_{i}\right)=i\left(\bigl|e_{i}\bigr\rangle\bigl\langle g_{i}\bigr|-\bigl|g_{i}\bigr\rangle\bigl\langle e_{i}\bigr|\right)=\begin{bmatrix}\begin{array}{cc}
0 & -i\\
i & 0
\end{array}\end{bmatrix}$. We also solve for $\rho(t)$ (solve Eq. (\ref{eq:master equation reduced}))
in that operator basis and the state basis of $\bigl|1\bigr\rangle=\bigl|g_{1}\bigr\rangle\bigl|g_{2}\bigr\rangle$,
$\bigl|2\bigr\rangle=\bigl|e_{1}\bigr\rangle\bigl|e_{2}\bigr\rangle$,
$\bigl|3\bigr\rangle=\bigl|g_{1}\bigr\rangle\bigl|e_{2}\bigr\rangle$,
and $\bigl|4\bigr\rangle=\bigl|e_{1}\bigr\rangle\bigl|g_{2}\bigr\rangle$,
where we assume that the initial state of the system is $\bigl|\Psi(0)\bigr\rangle=\bigl|e_{1}\bigr\rangle\bigl|g_{2}\bigr\rangle$,
i.e., only atom (qubit) one is in the excited state.

\section{DC current induced nonreciprocal graphene plasmon polaritons\label{sec:DC-current-induced}}

If we bias the graphene sheet as shown in Fig. \ref{fig:Model-for-single-interface},
we will induce a drift current (drift velocity) on the graphene surface.
In the presence of this drift the local surface conductivity for the
graphene becomes nonlocal \cite{Morgado2018,CorreasSerrano2019,HassaniGangaraj2022}.
This results in nonreciprocal graphene plasmon polaritons.
\begin{figure}[tp]
\begin{centering}
\hfill{}\subfloat[]{\centering{}%
\begin{minipage}[b][5cm][c]{6cm}%
\begin{center}
\includegraphics[width=6cm,keepaspectratio,height=5cm]{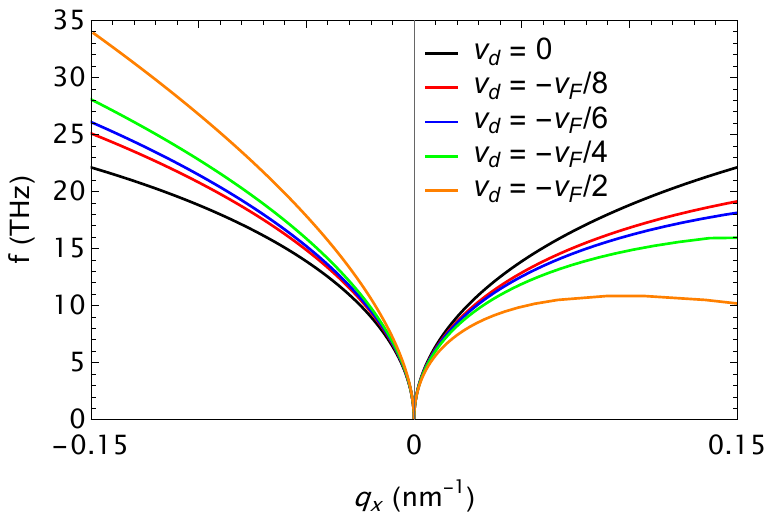}
\par\end{center}%
\end{minipage}}\hfill{}\subfloat[]{\begin{centering}
\includegraphics[width=6.2cm,keepaspectratio,height=5.2cm]{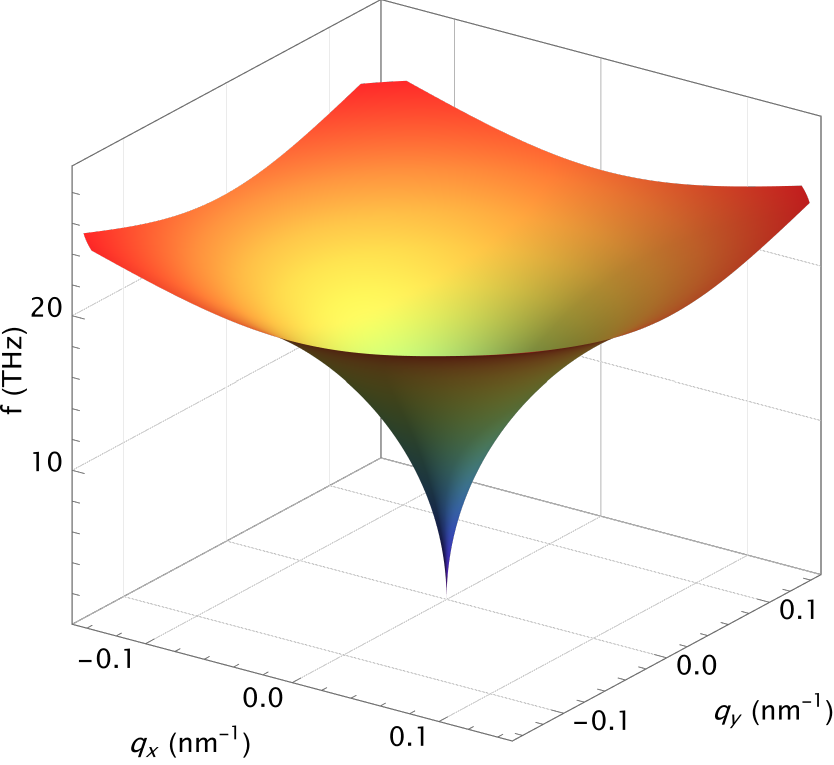}
\par\end{centering}
}\hfill{}
\par\end{centering}
\vspace{-0.6cm}

\begin{centering}
\hfill{}\subfloat[]{\begin{centering}
\includegraphics[width=6.2cm,keepaspectratio,height=5.2cm]{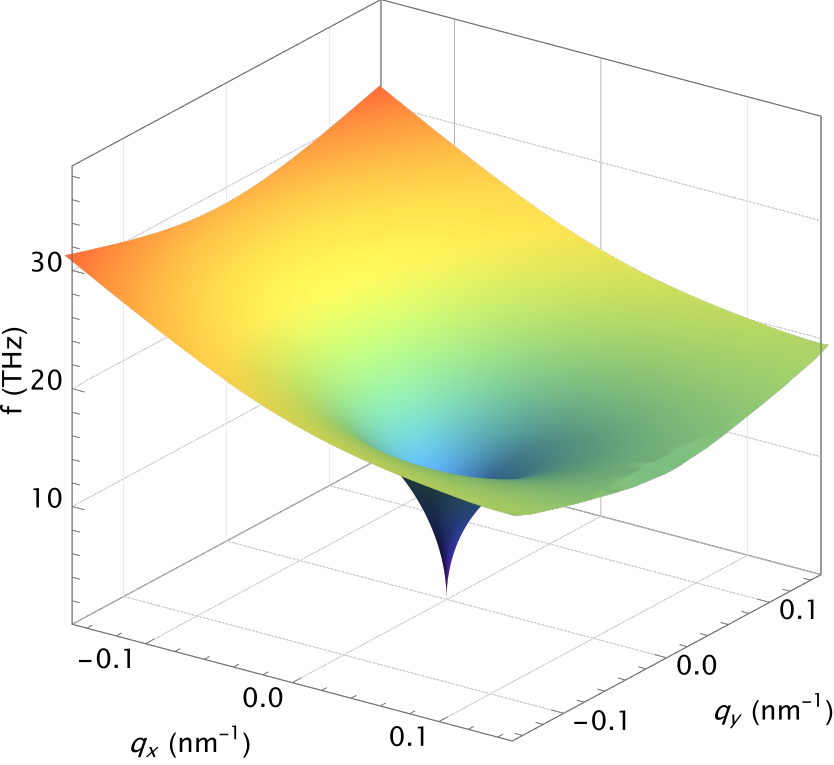}
\par\end{centering}
}\hfill{}\subfloat[]{\begin{centering}
\includegraphics[width=6.2cm,keepaspectratio,height=5.2cm]{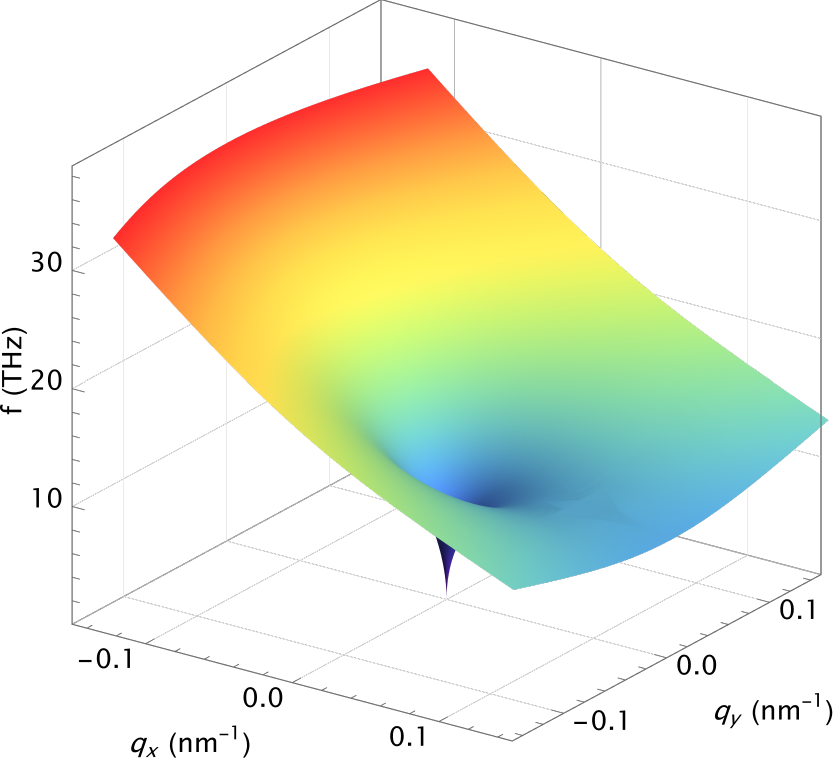}
\par\end{centering}
}\hfill{}
\par\end{centering}
\caption{Dispersion for TM surface waves supported by the graphene sheet for
(a) different drift velocity values, (b) graphene reciprocal (R) ($v_{d}=0$),
(c) graphene nonreciprocal (NR) for $v_{d}=\nicefrac{-v_{F}}{4}$,
and (d) graphene NR for $v_{d}=\nicefrac{-v_{F}}{2}$.\label{fig:Dispersion-for-TM}}
\end{figure}
\begin{figure}[tp]
\begin{centering}
\hfill{}\subfloat[]{\centering{}%
\begin{minipage}[b][3cm][c]{4cm}%
\begin{center}
\includegraphics[width=4cm,keepaspectratio,height=4cm]{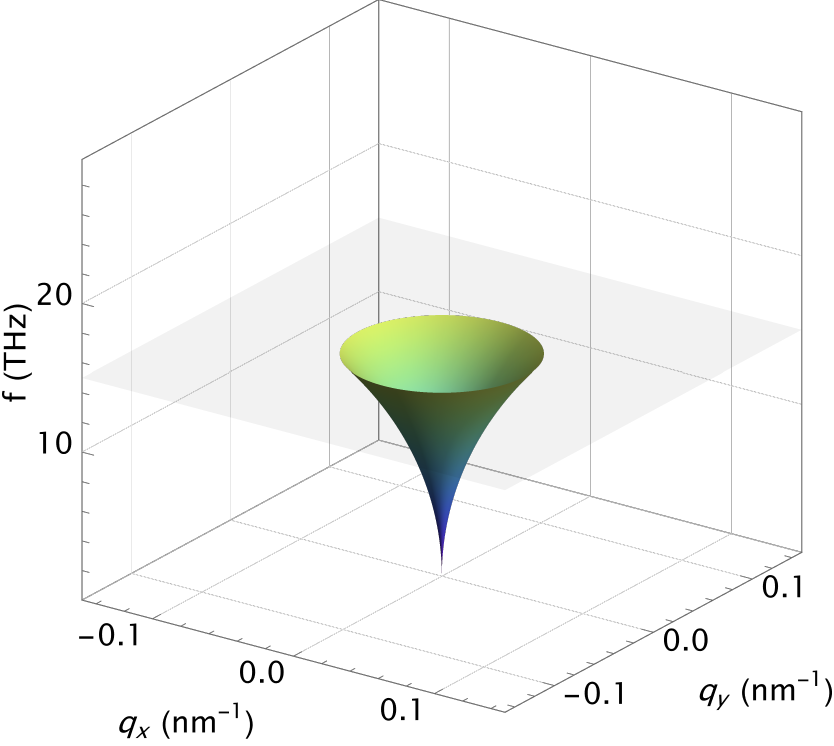}
\par\end{center}%
\end{minipage}}\hfill{}\subfloat[]{\centering{}\includegraphics[width=4cm,keepaspectratio,height=4cm]{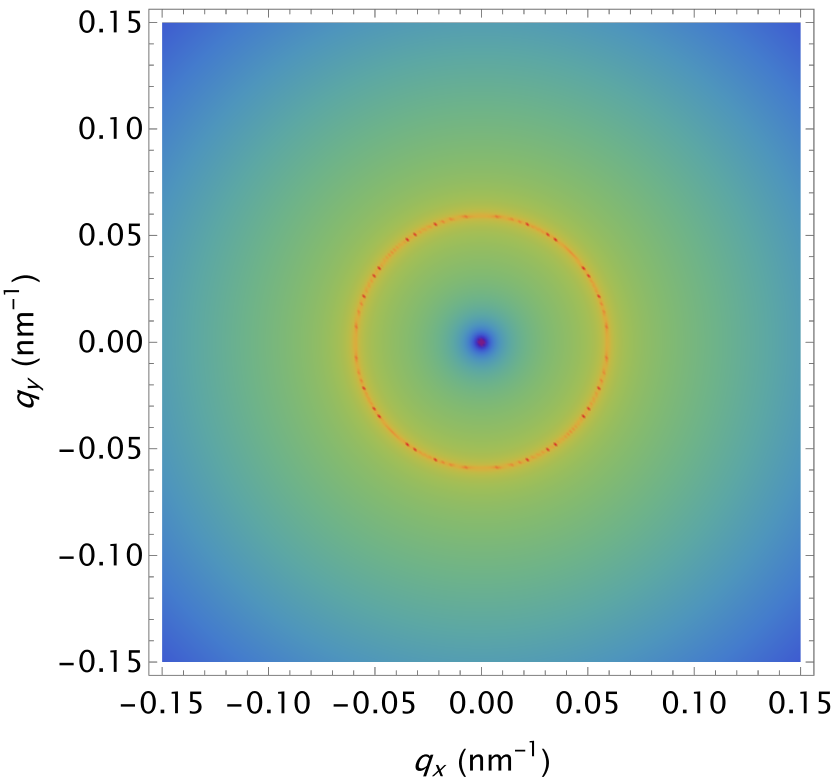}}\hfill{}\subfloat[]{\centering{}\includegraphics[width=4cm,keepaspectratio,height=4cm]{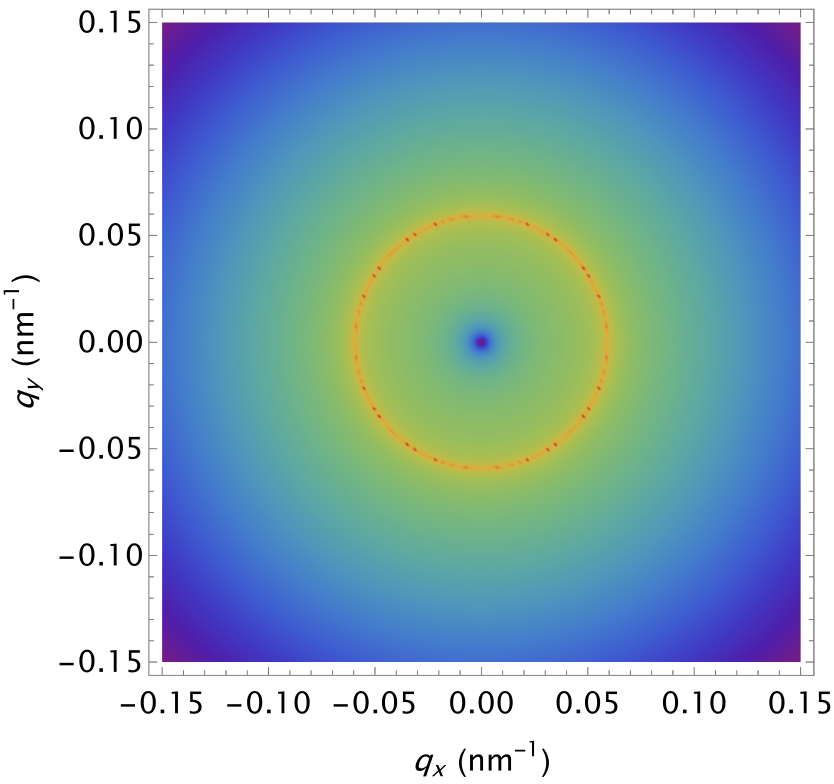}}\hfill{}
\par\end{centering}
\vspace{-0.8cm}

\begin{centering}
\hfill{}\subfloat[]{\begin{centering}
\begin{minipage}[b][3cm][c]{4cm}%
\begin{center}
\includegraphics[width=4cm,keepaspectratio,height=4cm]{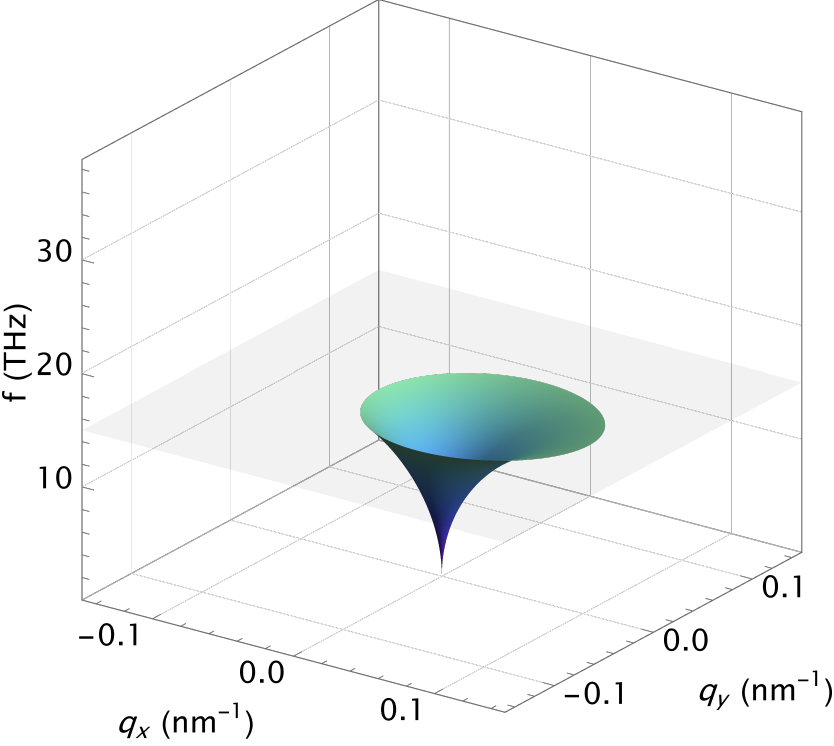}
\par\end{center}%
\end{minipage}
\par\end{centering}
\centering{}}\hfill{}\subfloat[]{\begin{centering}
\includegraphics[width=4cm,keepaspectratio,height=4cm]{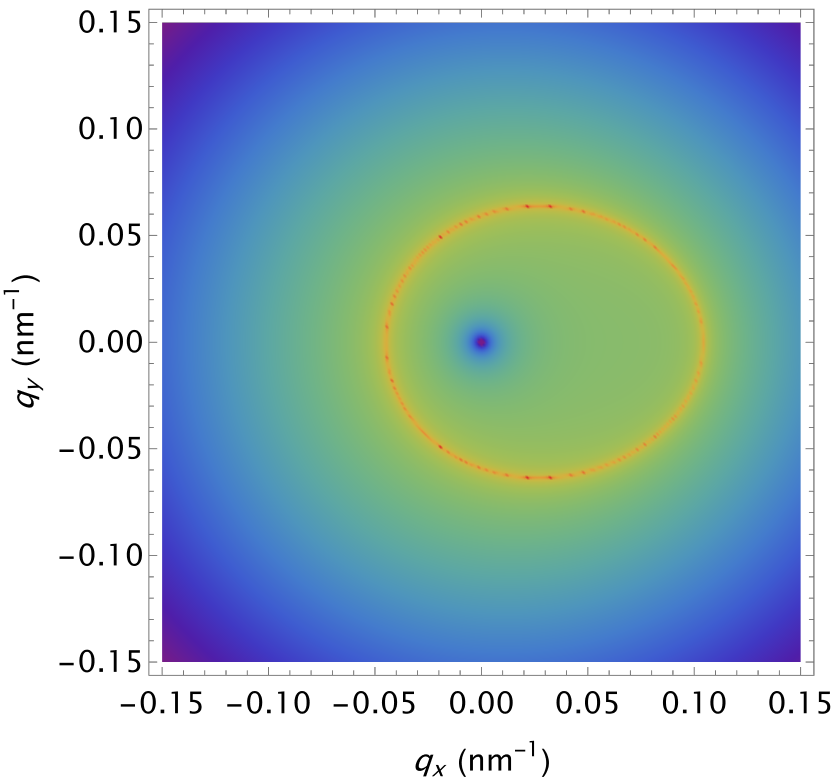}
\par\end{centering}
}\hfill{}\subfloat[]{\begin{centering}
\includegraphics[width=4cm,keepaspectratio,height=4cm]{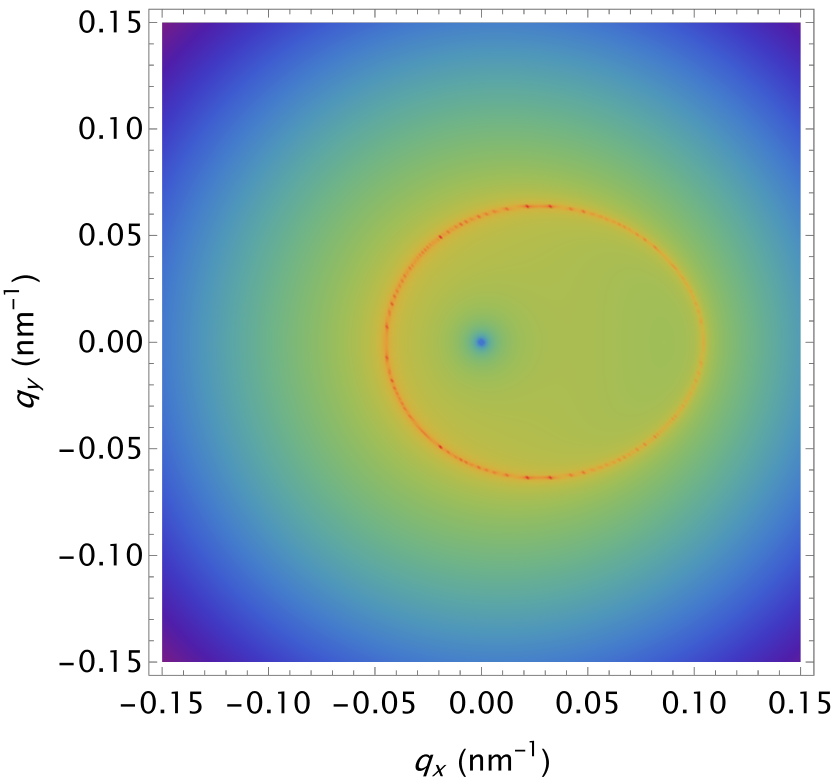}
\par\end{centering}
}\hfill{}
\par\end{centering}
\vspace{-0.8cm}

\begin{centering}
\hfill{}\subfloat[]{\begin{centering}
\begin{minipage}[b][3cm][c]{4cm}%
\begin{center}
\includegraphics[width=4cm,keepaspectratio,height=4cm]{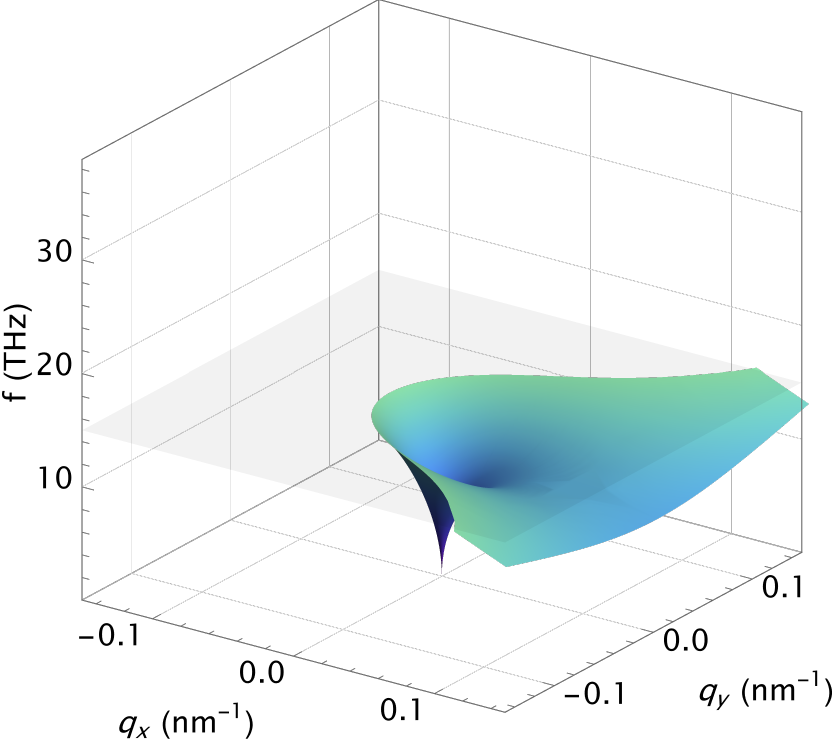}
\par\end{center}%
\end{minipage}
\par\end{centering}
}\hfill{}\subfloat[]{\begin{centering}
\includegraphics[width=4cm,keepaspectratio,height=4cm]{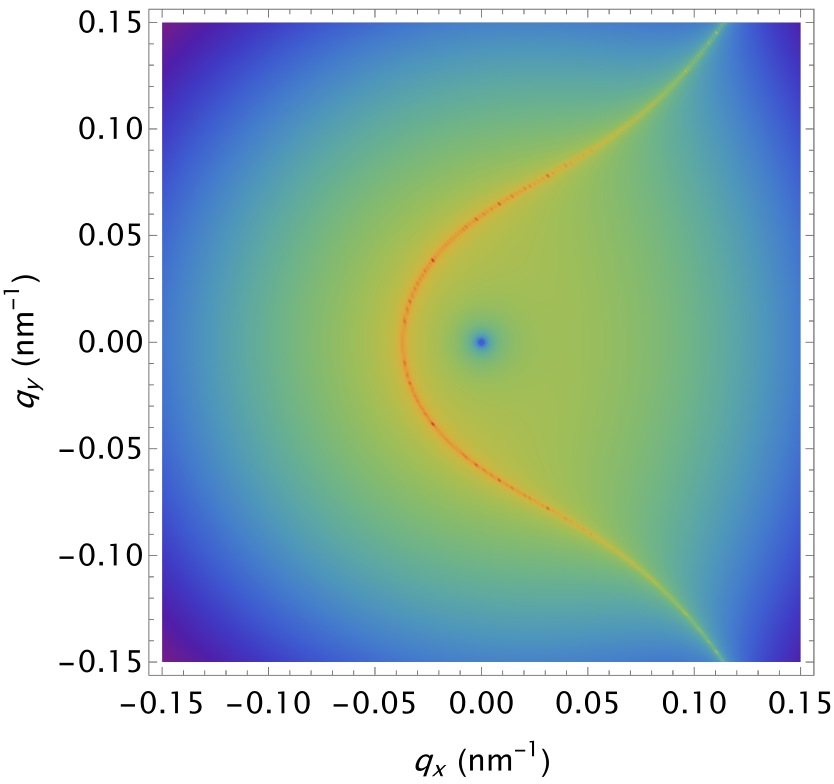}
\par\end{centering}
}\hfill{}\subfloat[]{\begin{centering}
\includegraphics[width=4cm,keepaspectratio,height=4cm]{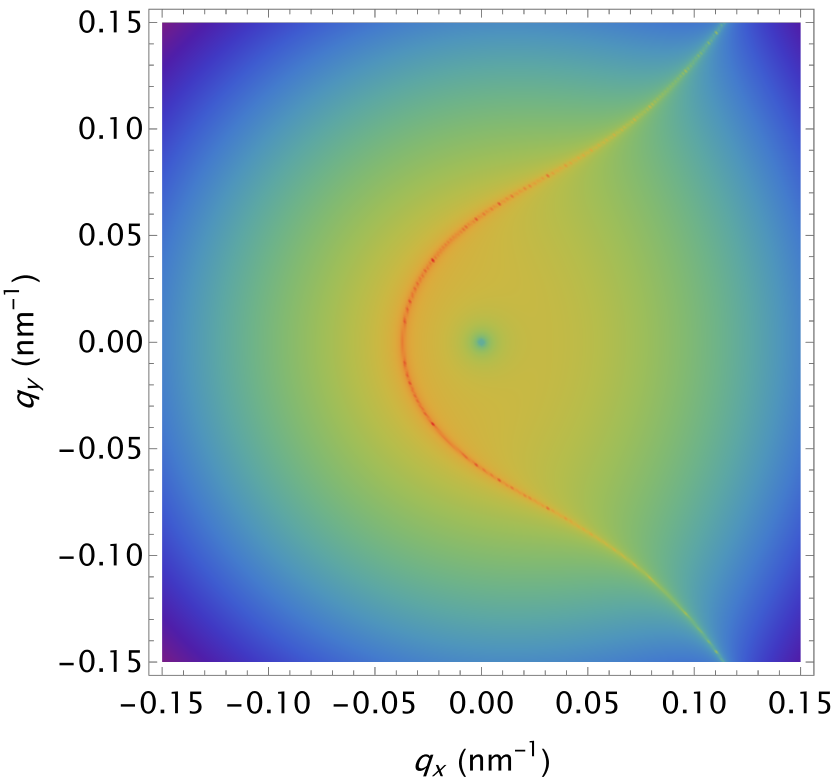}
\par\end{centering}
}\hfill{}
\par\end{centering}
\caption{Effect of the source height on the dispersion for TM surface waves
(the SPPs) supported by the graphene sheet: (a) graphene reciprocal
(R) ($v_{d}=0$) with cut-plane at 15 THz, (b) graphene R ($\lambda=\lambda_{pr}$)
with $z=z^{\prime}=\nicefrac{\lambda}{4}$, (c) graphene R ($\lambda=\lambda_{pr}$)
with $z=z^{\prime}=\nicefrac{\lambda}{3}$, (d) graphene nonreciprocal
(NR) for $v_{d}=\nicefrac{-v_{F}}{4}$ with cut-plane at 15 THz, (e)
graphene NR ($\lambda=\lambda_{pnr2}$) for $v_{d}=\nicefrac{-v_{F}}{4}$
with $z=z^{\prime}=\nicefrac{\lambda}{4}$, (f) graphene NR ($\lambda=\lambda_{pnr2}$)
for $v_{d}=\nicefrac{-v_{F}}{4}$ with $z=z^{\prime}=\nicefrac{\lambda}{3}$,
(g) graphene NR for $v_{d}=\nicefrac{-v_{F}}{2}$ with cut-plane at
15 THz, (h) graphene NR ($\lambda=\lambda_{pnr1}$) for $v_{d}=\nicefrac{-v_{F}}{2}$
with $z=z^{\prime}=\nicefrac{\lambda}{4}$, (i) graphene NR ($\lambda=\lambda_{pnr1}$)
for $v_{d}=\nicefrac{-v_{F}}{2}$ with $z=z^{\prime}=\nicefrac{\lambda}{3}$,
where the magnitude of the Green's function integrand is plotted (in
arb. units) for (b), (c), (e), (f), (h), and (i).\label{fig:Dispersion-for-TM-source height}}
\end{figure}

\subsection{Nonlocal surface conductivity for DC biased graphene\label{subsec:Nonlocal-Surface-Conductivity}}

We now define the nonlocal surface conductivity $\sigma_{d}$ for
DC biased graphene as \cite{Morgado2018,CorreasSerrano2019,HassaniGangaraj2022}
\begin{equation}
\sigma_{d}(v_{d},q_{x},\omega)=\frac{\omega}{\omega-q_{x}v_{d}}\sigma(\omega-q_{x}v_{d})=\frac{\omega}{\omega-\left|q\right|\cos(\phi)v_{d}}\sigma(\omega-\left|q\right|\cos(\phi)v_{d}),
\end{equation}
where $v_{d}$ is the drift velocity and $q_{x}$ is the wavenumber
in the momentum space along the $x$-direction on the graphene surface,
where we have assumed that the velocity distribution (e.g., Maxwell-Boltzmann)
is such that $v_{d}\approx v_{x}\gg v_{y}$. The resulting nonreciprocal
graphene plasmon polaritons become more unidirectional for higher
drift velocities, where, for graphene, given its high electrical conductivity
\cite{Neto2009,Shishir2009,Dorgan2010,Ozdemir2015,Ramamoorthy2016,Yamoah2017,Sabbaghi2018,Sabbaghi2022},
drift velocities on the order of $v_{F}\approx\nicefrac{c}{300}$
($v_{F}$ being the Fermi velocity) are possible \cite{Morgado2018}.

A plot for the nonlocal surface conductivity for graphene (in the
low-temperature limit) for $v_{d}=\nicefrac{-v_{F}}{2}$ is provided
in Fig. \ref{fig:Local-nonlocal-surface-conductivity}, where we can
see that the conductivity is nonreciprocal with respect to $q_{x}$
(larger conductivity in the direction of the drift velocity ($-q_{x}$-direction)),
e.g., see Fig. \ref{fig:nonlocal conductivity for 15 THz subfigure},
which is the nonlocal conductivity at 15 THz. In other words, the
conductivity is tilted along the $q_{x}$-axis; smaller in the $+q_{x}$-direction,
becoming larger in the $-q_{x}$-direction (the direction of the drift
velocity). This is commensurate with the nonreciprocal response for
the SPPs, where the smaller nonlocal $\mathrm{Im}\left(\sigma\right)$
values in the $+q_{x}$-direction are not sufficient for a strong
SPP response \cite{Liang2015}.
\begin{figure}[tp]
\begin{centering}
\includegraphics[width=6cm,keepaspectratio,height=5cm]{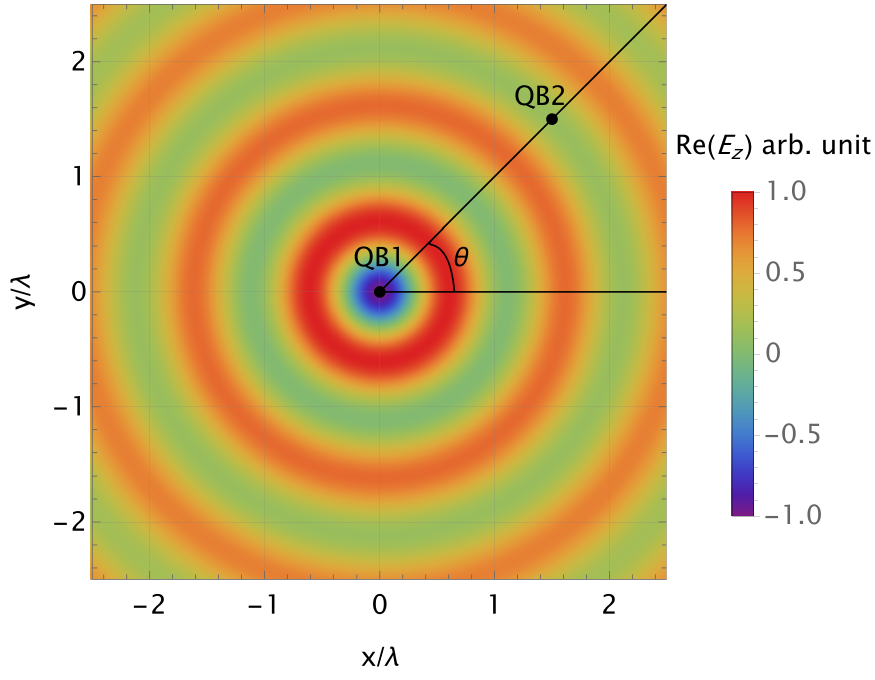}\includegraphics[width=6cm,keepaspectratio,height=5cm]{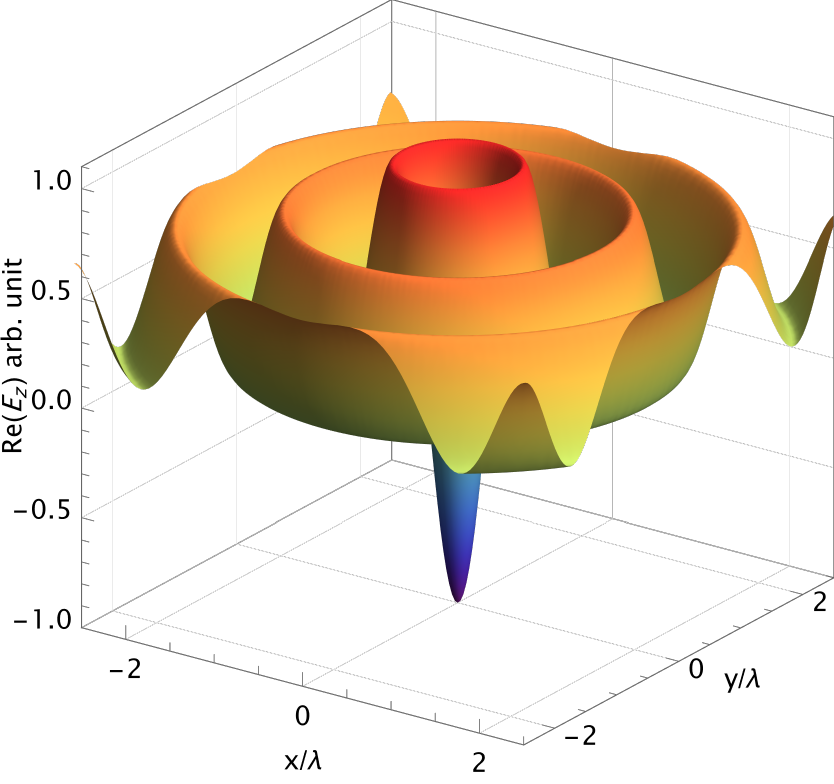}
\par\end{centering}
\caption{Electric field wave propagation for graphene R ($v_{d}=0$), ($\lambda=\lambda_{pr}$),
for $z=z^{\prime}=\nicefrac{\lambda}{3}$; the plot for $z=z^{\prime}=\nicefrac{\lambda}{4}$
(not provided) is similar.\label{fig:Electric-field-wave_graphene_R}}
\end{figure}
\begin{figure}[tp]
\begin{centering}
\hfill{}\subfloat[]{\centering{}\includegraphics[width=6cm,keepaspectratio,height=5cm]{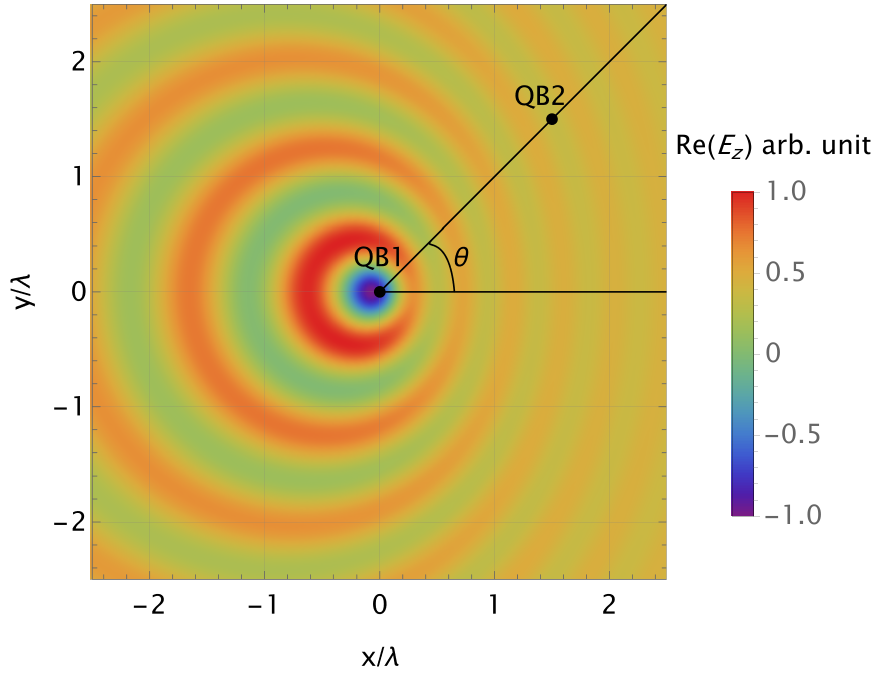}\includegraphics[width=6cm,keepaspectratio,height=5cm]{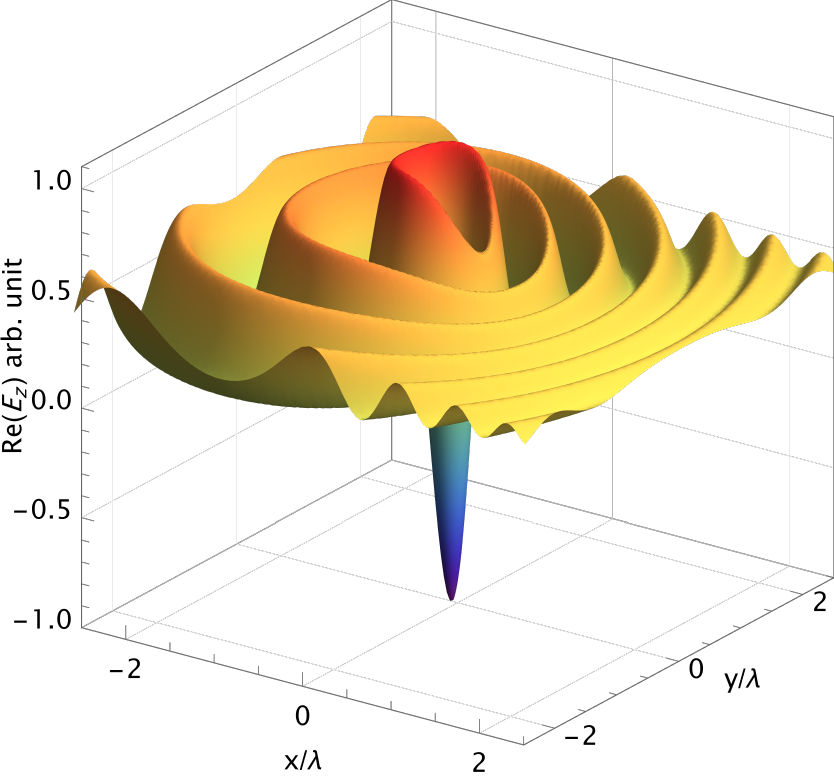}}\hfill{}
\par\end{centering}
\vspace{-0.6cm}

\begin{centering}
\hfill{}\subfloat[]{\begin{centering}
\includegraphics[width=6cm,keepaspectratio,height=5cm]{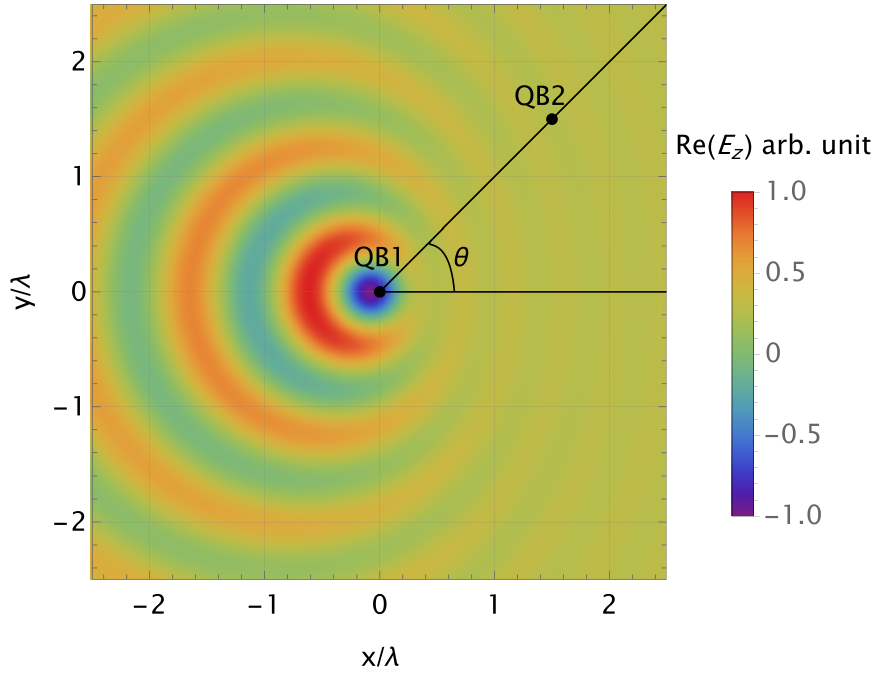}\includegraphics[width=6cm,keepaspectratio,height=5cm]{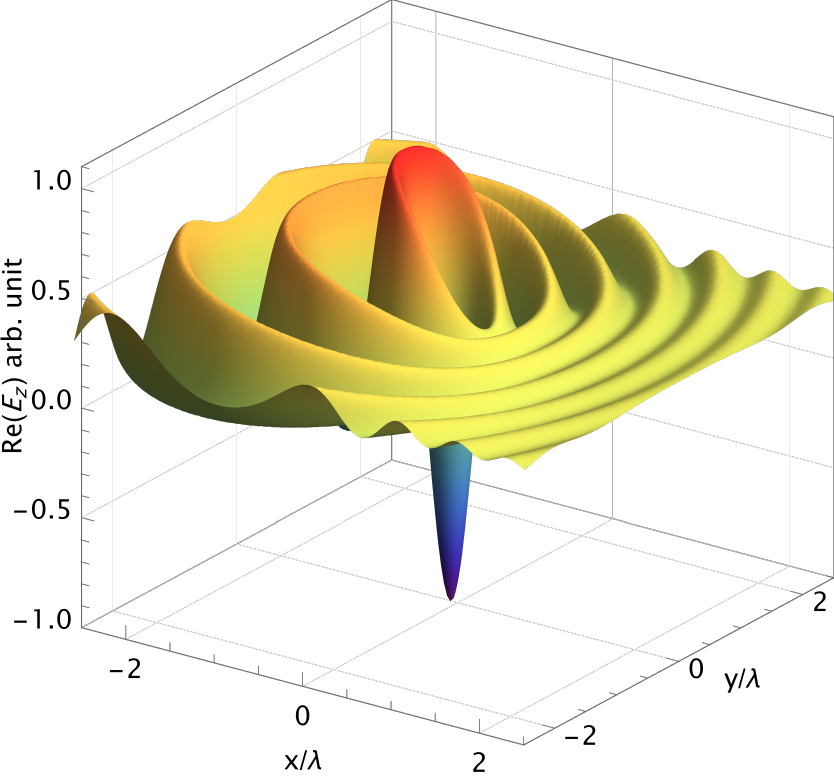}
\par\end{centering}
}\hfill{}
\par\end{centering}
\caption{Electric field wave propagation for graphene NR ($\lambda=\lambda_{pnr2}$)
for $v_{d}=\nicefrac{-v_{F}}{4}$: (a) $z=z^{\prime}=\nicefrac{\lambda}{4}$
and (b) $z=z^{\prime}=\nicefrac{\lambda}{3}$.\label{fig:Electric-field-wave_graphene_NR_vd_negvFdiv4}}
\end{figure}
\begin{figure}[tp]
\begin{centering}
\hfill{}\subfloat[]{\centering{}\includegraphics[width=6cm,keepaspectratio,height=5cm]{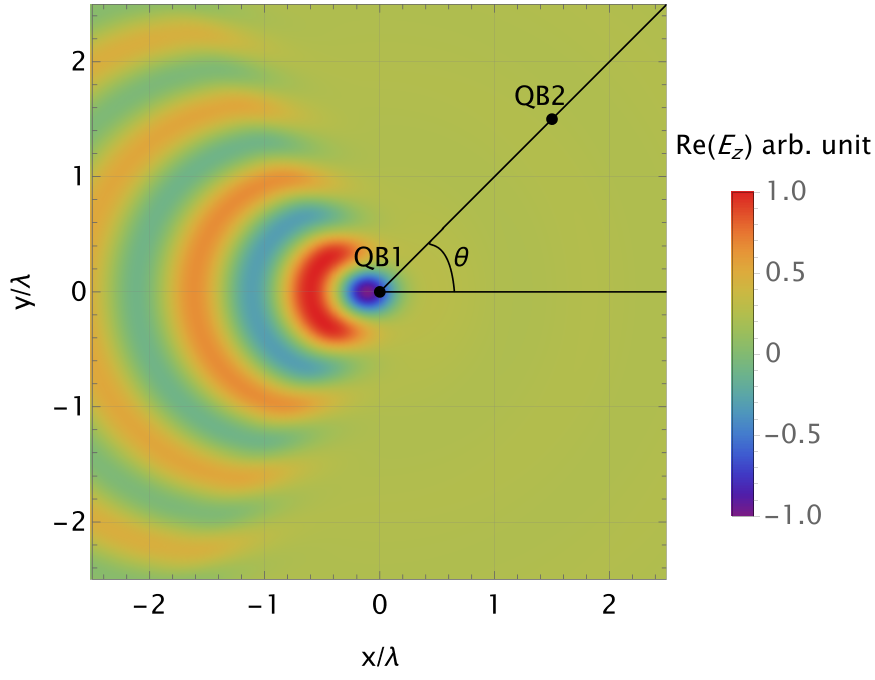}\includegraphics[width=6cm,keepaspectratio,height=5cm]{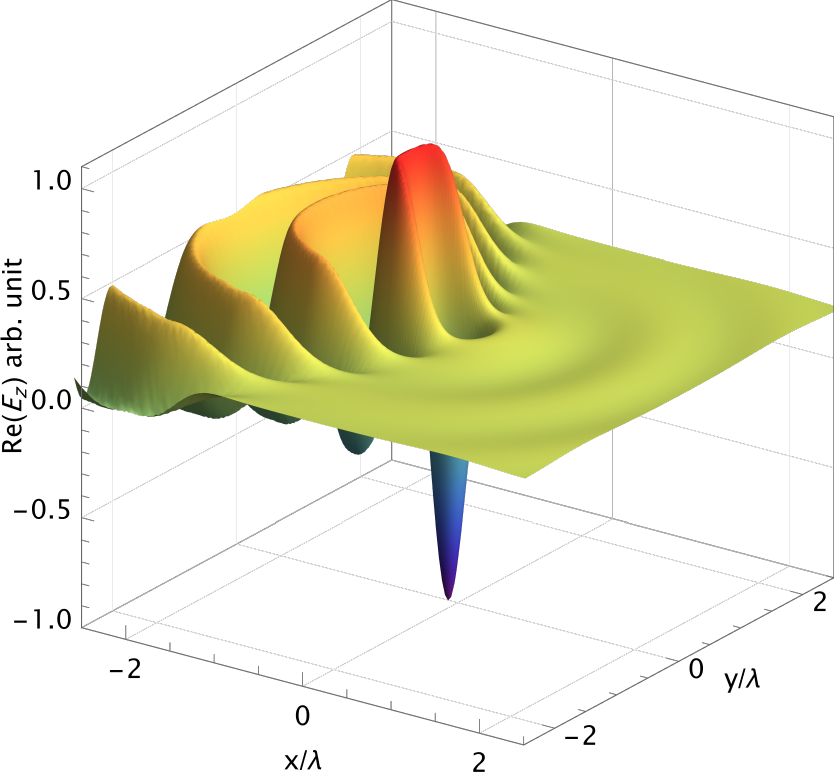}}\hfill{}
\par\end{centering}
\vspace{-0.6cm}

\begin{centering}
\hfill{}\subfloat[]{\begin{centering}
\includegraphics[width=6cm,keepaspectratio,height=5cm]{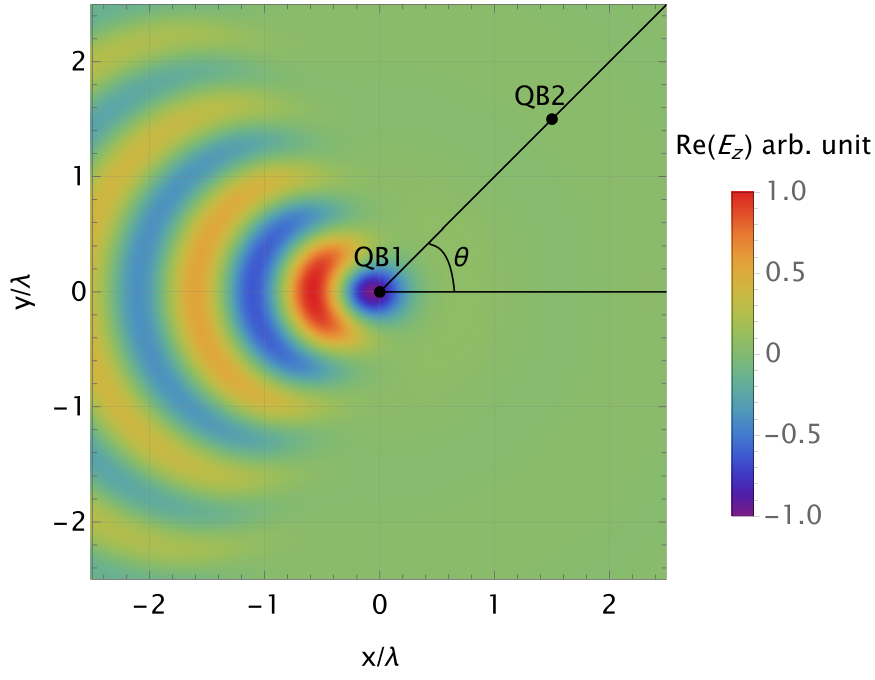}\includegraphics[width=6cm,keepaspectratio,height=5cm]{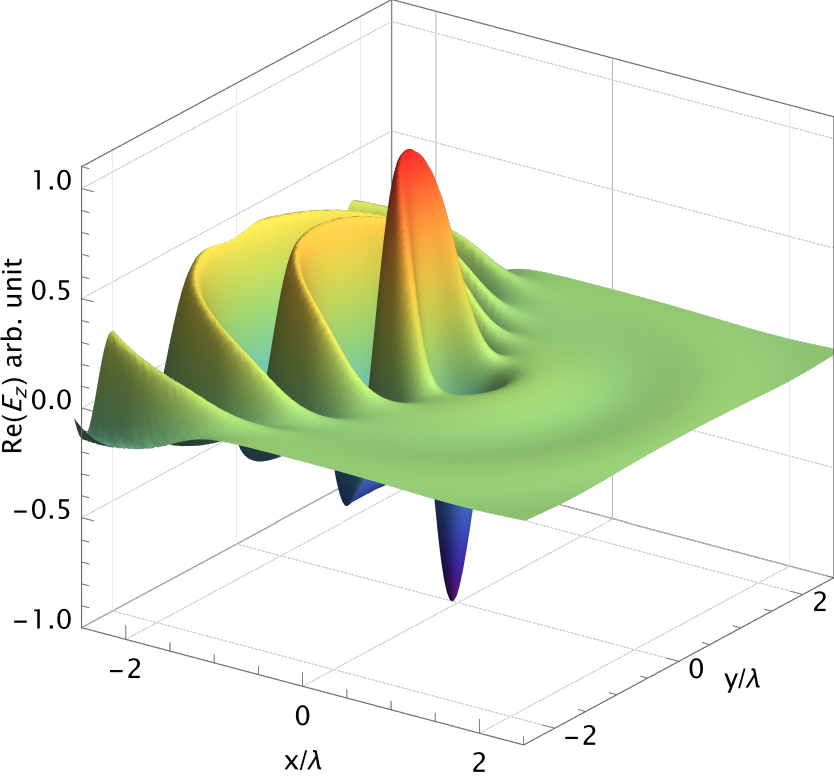}
\par\end{centering}
}\hfill{}
\par\end{centering}
\caption{Electric field wave propagation for graphene NR ($\lambda=\lambda_{pnr1}$)
for $v_{d}=\nicefrac{-v_{F}}{2}$: (a) $z=z^{\prime}=\nicefrac{\lambda}{4}$
and (b) $z=z^{\prime}=\nicefrac{\lambda}{3}$.\label{fig:Electric-field-wave_graphene_NR_vd_negvFdiv2}}
\end{figure}

\subsection{Dispersion\label{subsec:Dispersion}}

The dispersion relation for the TM surface waves (the SPPs) supported
by the graphene sheet is obtained by setting the denominator, $Z^{E}$,
in Eq. (\ref{eq:scattering coefficient}) equal to zero, which results
in
\begin{equation}
Z^{E}=\frac{\varepsilon_{1}}{\varepsilon_{2}}p_{2}+p_{1}+\frac{\sigma_{d}p_{2}p_{1}}{-i\omega\varepsilon_{2}}=0.\label{eq:dispersion relation}
\end{equation}
Then to determine the dispersion we obtain the resulting solution
of Eq. (\ref{eq:dispersion relation}). Note that throughout this
work, when obtaining the results pertaining to the Green's function
values (and the corresponding results) the applicable configuration
is the graphene sheet embedded in $\mathrm{SiO_{2}}$, i.e., the relative
permittivity in both regions (region 1 and region 2) is $\varepsilon_{r1}=\varepsilon_{r2}=4$.

The dispersion for different drift velocity values is shown in Fig.
\ref{fig:Dispersion-for-TM}, where we can see that for no drift velocity
the dispersion is reciprocal for all frequencies, however, for even
a small amount of drift velocity there is a nonreciprocal response.
For the larger drift velocity values propagation becomes unidirectional
(in the direction of the drift velocity), where the SPPs only propagate
in one direction above certain frequencies, starting at fairly low
THz frequencies for the larger drift velocity values.

We can also see the effect of the source height on the SPP dispersion
by looking at the magnitude of the Green's function integrand for
the Sommerfeld integrals at an equi-frequency contour (EFC) at 15
THz. From the plots in Fig. \ref{fig:Dispersion-for-TM-source height}
we can see that the intensity of the SPPs (the poles) in the Green's
function integrand varies along the EFC as the source height changes;
different parts of the EFC become dominant and contribute more as
the source height changes.

Throughout this work the qubit separation distances and source (observation)
heights are normalized to wavelength. Given the disparity between
the wavelengths (at 15 THz the vacuum wavelength is approx. 117 times
larger than the SPP wavelengths), we normalize to each respective
wavelength, i.e., the qubit separation distances and source (observation)
heights are with respect to the 'electrical lengths.' The normalization
wavelengths used for vacuum, graphene reciprocal (R), graphene nonreciprocal
(NR) for $v_{d}=\nicefrac{-v_{F}}{2}$, and graphene nonreciprocal
(NR) for $v_{d}=\nicefrac{-v_{F}}{4}$ are $\lambda_{0}\approx19.986\:\mathrm{\mu m}$,
$\lambda_{pr}\approx0.106\:\mathrm{\mu m}$, $\lambda_{pnr1}\approx0.171\:\mathrm{\mu m}$,
and $\lambda_{pnr2}\approx0.137\:\mathrm{\mu m}$, respectively.

For no drift velocity the dispersion is reciprocal, and the SPPs (the
poles) contribute uniformly to the Green's function integrand along
the EFC resulting in the SPP propagation being reciprocal. For nonzero
drift velocities the dispersion is nonreciprocal, and the SPPs (the
poles) contribute nonuniformly, where there are more poles contributing
at different parts of the EFC, which affects the shape and direction
of the SPPs propagating on the graphene sheet. 

For shorter source heights the SPPs at the extents of the EFC contribute;
the shape and direction of the SPP propagation is commensurate. Finally,
for larger source heights the SPPs at the extents of the EFC contribute
less, and those towards the center are more dominant; the shape and
direction of the SPP propagation is more focused in the direction
of the drift velocity.

\subsection{Electric field}

The photon model described here is fully (macroscopically) quantum,
although the Green's function provides the classical electric field,
which is useful to envision the surface plasmons. In order to see
the wave propagation for the reciprocal and nonreciprocal cases we
plot the classical electric field, where we can also see the effect
of the source height on the electric field response. We obtain the
electric field values from the Green's function values using Eq. (\ref{eq:electric field})
(all results are for the frequency set to 15 THz). The corresponding
plots are in Figs. \ref{fig:Electric-field-wave_graphene_R}, \ref{fig:Electric-field-wave_graphene_NR_vd_negvFdiv4},
and \ref{fig:Electric-field-wave_graphene_NR_vd_negvFdiv2}.

For graphene R the wave propagation is the same in all directions,
as expected. In the case of graphene NR the wave propagation becomes
unidirectional for larger drift velocity values, in the direction
of the drift velocity. We also see that for larger source heights,
that are still in the vicinity of the interface, the shape and direction
of the propagation is more focused in the direction of the drift velocity,
which is commensurate with what was observed for the effect of the
source height on the dispersion for the SPPs. However, if the source
height becomes too large, i.e., too far from the interface, the source
will not couple with the interface enough to result in a strong nonreciprocal
SPP response.
\begin{figure}[tp]
\begin{centering}
\hfill{}\subfloat[\label{fig:concurrence vs angle subfigure}]{\centering{}\includegraphics[width=8cm,keepaspectratio,height=6cm]{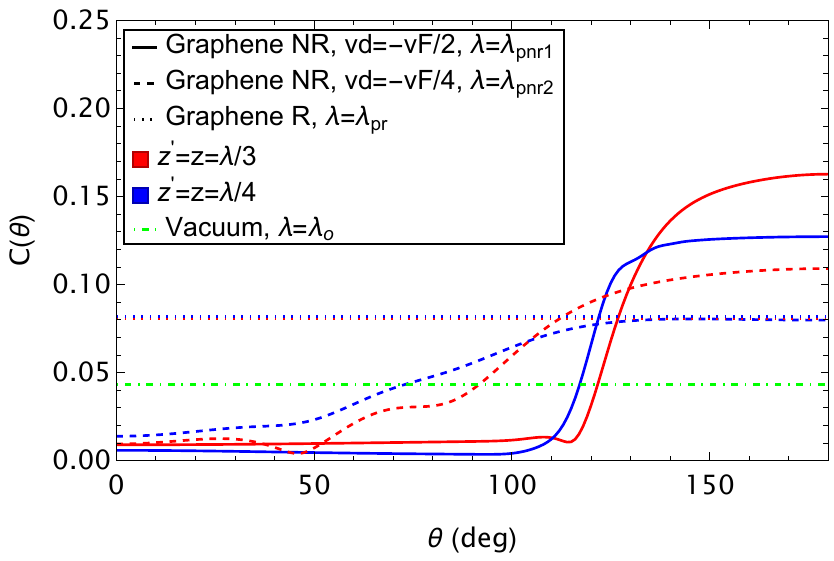}}\hfill{}
\par\end{centering}
\vspace{-0.5cm}

\begin{centering}
\hfill{}\subfloat[\label{fig:concurrence vs rho subfigure}]{\begin{centering}
\includegraphics[width=8cm,keepaspectratio,height=6cm]{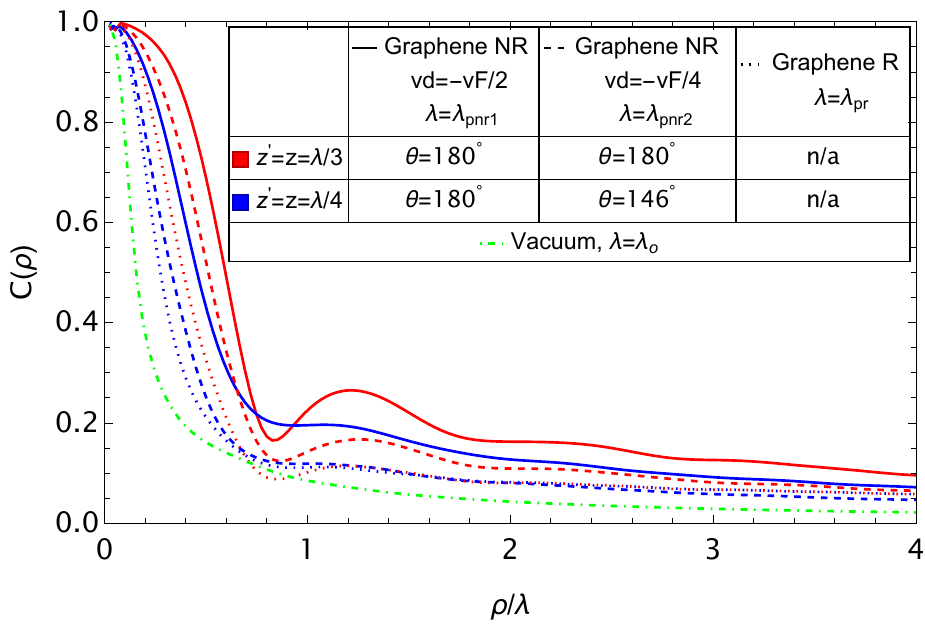}
\par\end{centering}
}\hfill{}
\par\end{centering}
\caption{Concurrence versus angle and qubit separation distance for the qubit
position configuration defined in Figs. \ref{fig:Electric-field-wave_graphene_R},
\ref{fig:Electric-field-wave_graphene_NR_vd_negvFdiv4}, and \ref{fig:Electric-field-wave_graphene_NR_vd_negvFdiv2},
where we consider $0\lyxmathsym{\protect\textdegree}\protect\leq\theta\protect\leq180\lyxmathsym{\protect\textdegree}$
only given the symmetry about the $y$-axis: (a) concurrence versus
angle, where we obtained the maximum $C(t)$, at a qubit separation
distance of $\rho=2\lambda$, for each angle, i.e., $C(\theta)=\mathrm{max}_{t}\left(C(\rho,\theta,t)\right)$,
and (b) concurrence versus qubit separation distance, where we obtained
the maximum $C(t)$, at the maximum angle determined in (a), where
applicable, for each qubit separation distance $\rho$, i.e., $C(\rho)=\mathrm{max}_{t}\left(C(\rho,\theta,t)\right)$.
In all cases the plots were done for two different observation and
source heights, $z=z^{\prime}=\nicefrac{\lambda}{4}$ and $z=z^{\prime}=\nicefrac{\lambda}{3}$.
\label{fig:Concurrence-versus-angle}}
\end{figure}
\begin{figure}[tp]
\begin{centering}
\hfill{}\subfloat[\label{fig:concurrences vs transient time_subfigure}]{\centering{}\includegraphics[width=8cm,keepaspectratio,height=6cm]{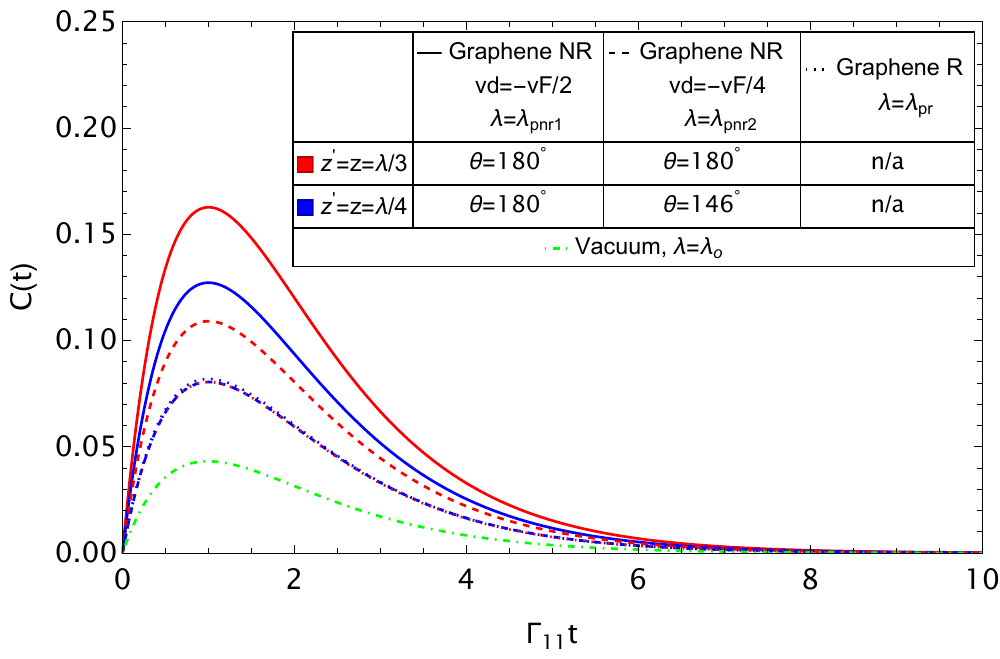}}\hfill{}
\par\end{centering}
\vspace{-0.5cm}

\begin{centering}
\hfill{}\subfloat[\label{fig:concurrences vs steady state time_subfigure}]{\begin{centering}
\includegraphics[width=8cm,keepaspectratio,height=6cm]{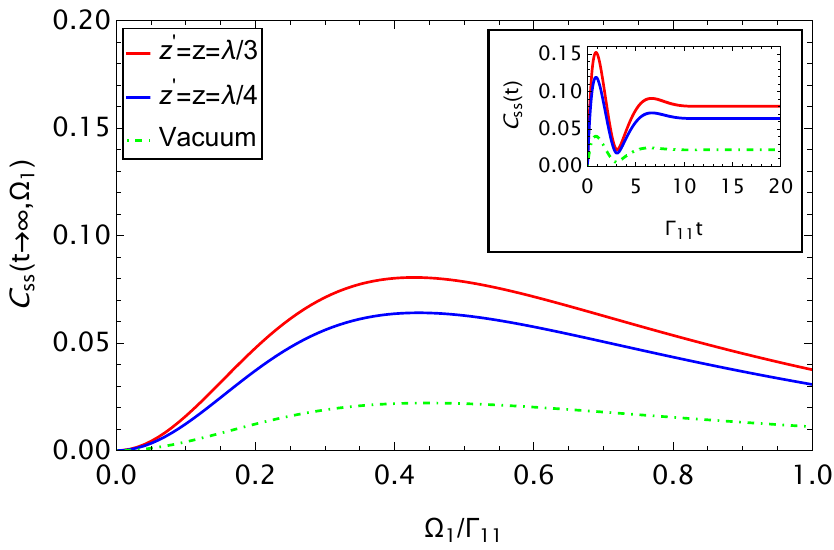}
\par\end{centering}
}\hfill{}
\par\end{centering}
\caption{Concurrence versus time, where we look at $C(t)$ for the maximum
angle determined in Fig. \ref{fig:concurrence vs angle subfigure}:
(a) concurrence versus transient time (the pump intensities $\mathrm{\varOmega}_{1}=\mathrm{\varOmega}_{2}=0$)),
$C(t)$, at the maximum angle and a qubit separation distance of $\rho=2\lambda$,
i.e., $C(t)=\mathrm{max}_{\theta}\left(C(\rho,\theta,t)\right)$,
and (b) steady state concurrence at $t\rightarrow\infty$, $C_{ss}(t\rightarrow\infty)$,
versus $\mathrm{\varOmega}_{1}$, for the maximum angle and a qubit
separation distance of $\rho=2\lambda$, i.e., $C_{ss}(t\rightarrow\infty,\mathrm{\varOmega}_{1})=\mathrm{max}_{\theta}\left(C(\rho,\theta,t\rightarrow\infty,\mathrm{\varOmega}_{1})\right)$,
where the laser pump intensity at QB2 is set to zero ($\mathrm{\varOmega}_{2}=0$).
The inset in (b) is $C_{ss}(t)$ for the $\mathrm{\varOmega}_{1}$
value where $C_{ss}(t\rightarrow\infty)$ is maximum ($\mathrm{\varOmega}_{1}=0.45\varGamma_{11}$
for the vacuum plot and $\mathrm{\varOmega}_{1}=0.43\varGamma_{11}$
for the other plots), where ($\mathrm{\varOmega}_{2}=0$). In (b)
the plots were done for the cases of $v_{d}=\nicefrac{-v_{F}}{2}$
and vacuum, and in both (a) and (b), the plots for $v_{d}=\nicefrac{-v_{F}}{2}$
were done for two different observation and source heights, $z=z^{\prime}=\nicefrac{\lambda}{4}$
and $z=z^{\prime}=\nicefrac{\lambda}{3}$. \label{fig:Concurrence-versus-time}}
\end{figure}

\section{Results\label{sec:Results}}

To determine which direction concurrence is maximized we plot concurrence
versus angle, where we consider $0\lyxmathsym{\textdegree}\leq\theta\leq180\lyxmathsym{\textdegree}$
only given the symmetry about the $y$-axis. We position the qubits
in the configuration defined in Figs. \ref{fig:Electric-field-wave_graphene_R},
\ref{fig:Electric-field-wave_graphene_NR_vd_negvFdiv4}, and \ref{fig:Electric-field-wave_graphene_NR_vd_negvFdiv2},
where we set the qubit separation distance $\rho=2\lambda$ and sweep
$\theta$ from $0\lyxmathsym{\textdegree}$ to $180\lyxmathsym{\textdegree}$,
determining the maximum concurrence versus time at each angle, i.e.,
$C(\theta)=\mathrm{max}_{t}\left(C(\rho,\theta,t)\right)$. As seen
in Fig. \ref{fig:concurrence vs angle subfigure}, there is good control
over concurrence (entanglement) as a function of angle for the nonreciprocal
(NR) cases. Additionally, the concurrence is higher for larger drift
velocities, i.e., for higher directionality of the field. Also, in
the NR case, the concurrence is higher, at the maximum angle, for
the larger source height. Finally, there is good enhancement of entanglement
for the NR case, at the maximum angle, over the reciprocal (R) case
and vacuum.
\begin{figure}[tp]
\begin{centering}
\hfill{}\subfloat[\label{fig:entanglement control, vd=00003D-vF/2, subfigure}]{\centering{}%
\begin{minipage}[b][4cm][c]{4cm}%
\begin{center}
\includegraphics[width=4cm,keepaspectratio,height=4cm]{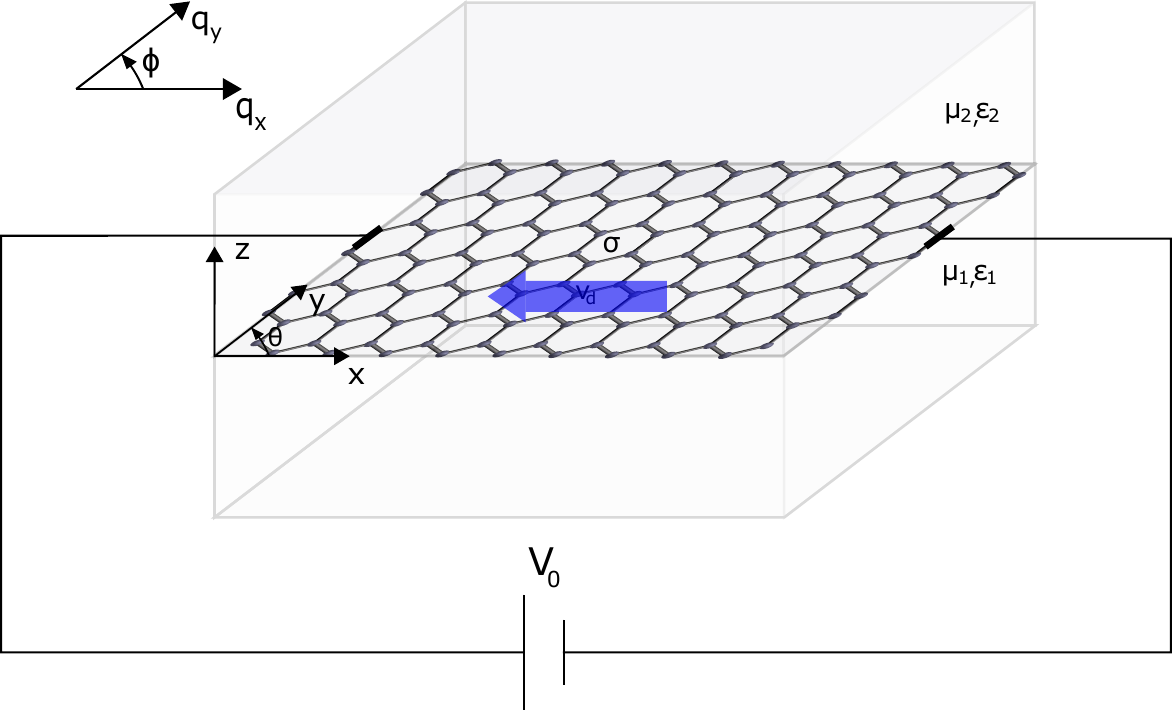}
\par\end{center}%
\end{minipage}\enskip{}\includegraphics[width=4cm,keepaspectratio,height=4cm]{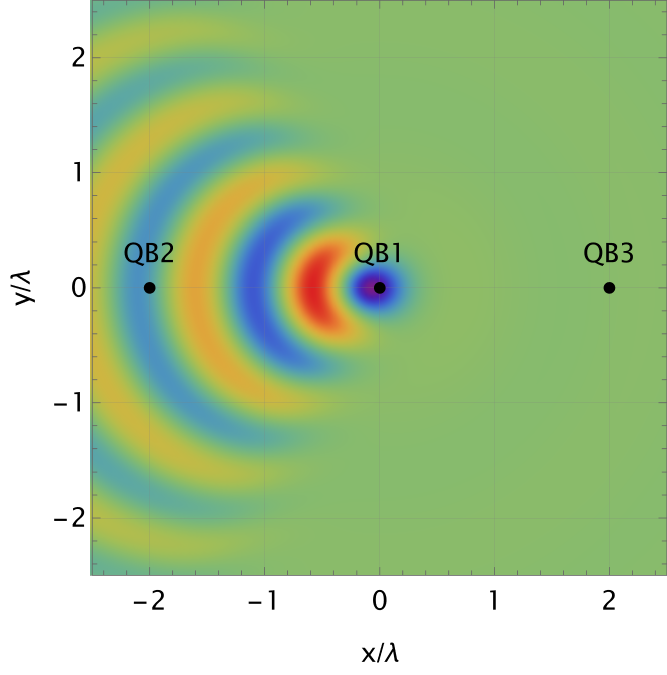}\enskip{}%
\begin{minipage}[b][4cm][c]{4.5cm}%
\begin{center}
\includegraphics[width=4.5cm,keepaspectratio,height=4cm]{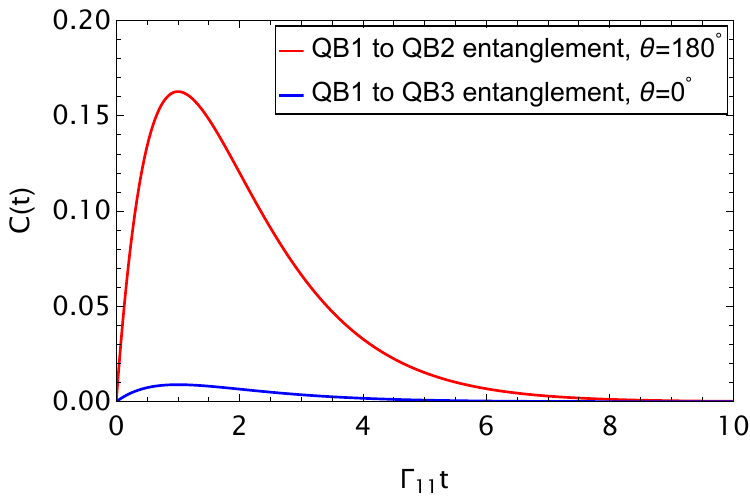}
\par\end{center}%
\end{minipage}}\hfill{}
\par\end{centering}
\vspace{-0.5cm}

\begin{centering}
\hfill{}\subfloat[\label{fig:entanglement control, vd=00003DvF/2, subfigure}]{\begin{centering}
\begin{minipage}[b][4cm][c]{4cm}%
\begin{center}
\includegraphics[width=4cm,keepaspectratio,height=4cm]{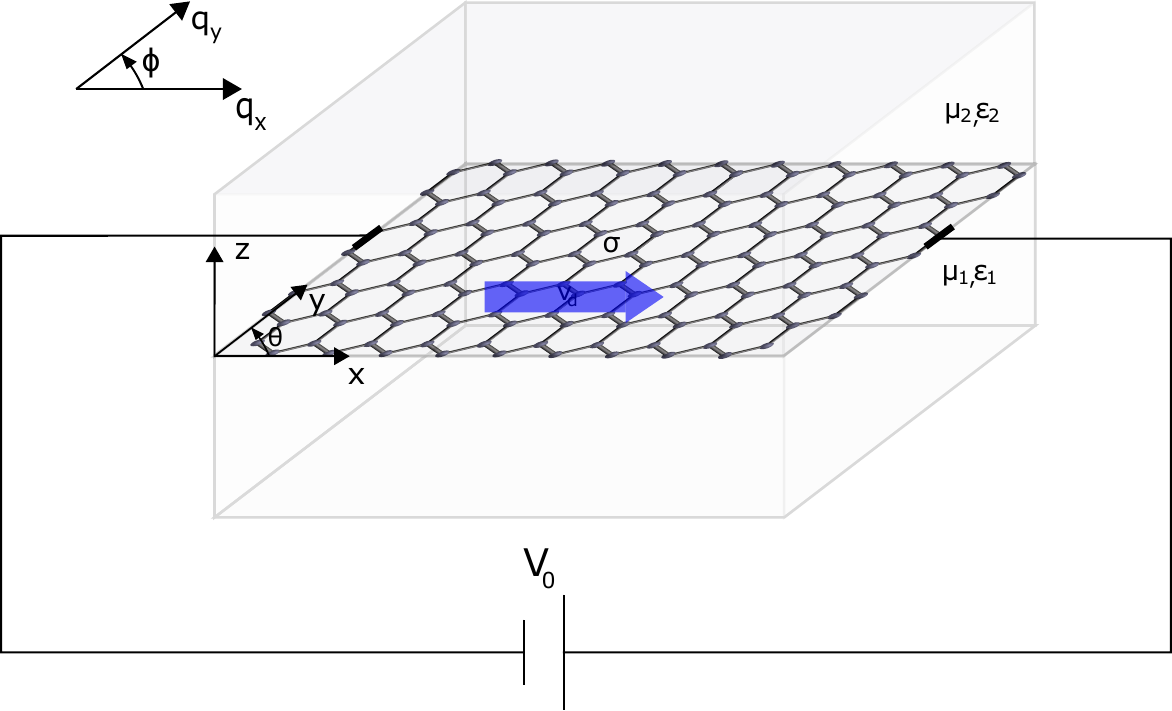}
\par\end{center}%
\end{minipage}\enskip{}\includegraphics[width=4cm,keepaspectratio,height=4cm]{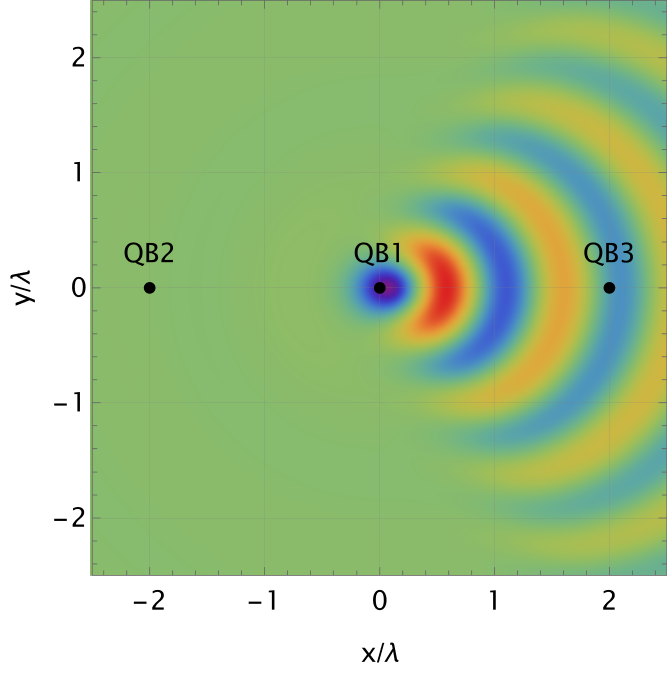}\enskip{}%
\begin{minipage}[b][4cm][c]{4.5cm}%
\begin{center}
\includegraphics[width=4.5cm,keepaspectratio,height=4cm]{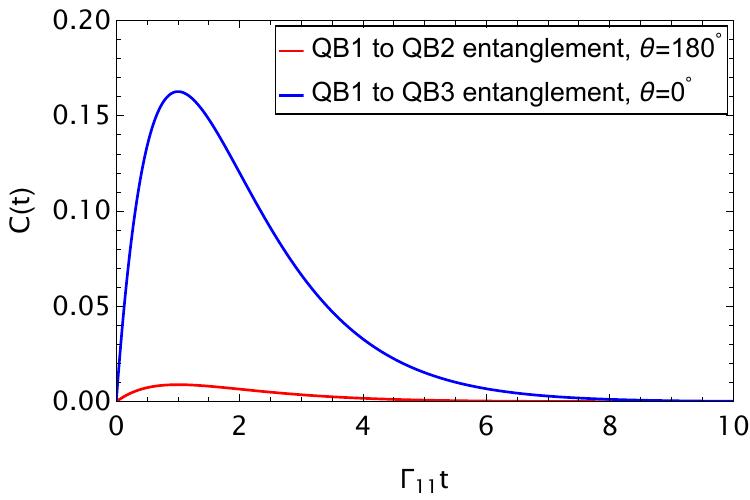}
\par\end{center}%
\end{minipage}
\par\end{centering}
}\hfill{}
\par\end{centering}
\caption{Configuration for use of the proposed system as a means for entanglement
control (note that for the proposed system configuration only pairwise
entanglement needs to be considered since coupling with the third
qubit is so weak that we can assume it doesn\textquoteright t exist):
(a) graphene biased such that $v_{d}=\nicefrac{-v_{F}}{2}$ and (b)
graphene biased such that $v_{d}=\nicefrac{v_{F}}{2}$. In all cases,
the observation and source heights are $z=z^{\prime}=\nicefrac{\lambda}{3}$
and the qubit separation distances are $\rho=2\lambda$. \label{fig:Configuration-for-use, entanglement control}}
\end{figure}

We then plot concurrence as a function of $\rho$, at the applicable
maximum angle, to see how the concurrences compare to each other with
respect to the qubit separation distance; determining the maximum
concurrence versus time at each $\rho$ value, i.e., $C(\rho)=\mathrm{max}_{t}\left(C(\rho,\theta,t)\right)$.
We can see from Fig. \ref{fig:concurrence vs rho subfigure} that
the entanglement for the NR case is better than the R case, and vacuum,
for fairly large qubit separation distances.

We also plot concurrence versus time (for the transient case, (the
pump intensities $\mathrm{\varOmega}_{1}=\mathrm{\varOmega}_{2}=0$)),
at the maximum angle and a qubit separation distance of $\rho=2\lambda$,
i.e., $C(t)=\mathrm{max}_{\theta}\left(C(\rho,\theta,t)\right)$,
to compare the concurrences. As seen in Fig. \ref{fig:concurrences vs transient time_subfigure},
there is good enhancement for the NR case over the R case and vacuum,
which is the case for larger qubit separation distances as well.

In order to maintain the entanglement an external coherent drive (i.e.,
a laser pump) is implemented at each qubit, where different pumping
profiles can be applied. We apply the laser pump at QB1 only, without
applying one at QB2 (the laser pump intensity at QB2 is set to zero
($\mathrm{\varOmega}_{2}=0$)). We determine the maximum steady state
concurrence at $t\rightarrow\infty$, $C_{ss}(t\rightarrow\infty)$,
by plotting $C_{ss}(t\rightarrow\infty)$ versus $\mathrm{\varOmega}_{1}$,
for the maximum angle and a qubit separation distance of $\rho=2\lambda$,
i.e., $C_{ss}(t\rightarrow\infty,\mathrm{\varOmega}_{1})=\mathrm{max}_{\theta}\left(C(\rho,\theta,t\rightarrow\infty,\mathrm{\varOmega}_{1})\right)$.
We then plot $C_{ss}(t)$ for the $\mathrm{\varOmega}_{1}$ value
where $C_{ss}(t\rightarrow\infty)$ is maximum (see the inset in Fig.
\ref{fig:concurrences vs steady state time_subfigure}). All of the
plots in Fig. \ref{fig:concurrences vs steady state time_subfigure}
are done for the cases of $v_{d}=\nicefrac{-v_{F}}{2}$ and vacuum.
As seen in Fig. \ref{fig:concurrences vs steady state time_subfigure},
$C_{ss}(t\rightarrow\infty,\mathrm{\varOmega}_{1})$ is small for
low and high values for $\mathrm{\varOmega}_{1}$, and is maximum
for values that are close to the middle of that range.

Since it was shown that there is good control over concurrence (entanglement)
as a function of angle, we can use this configuration to control entanglement
(controlling which qubits are entangled by means of changing the polarity
of the DC bias, i.e., the direction of the drift velocity). As seen
in Fig. \ref{fig:entanglement control, vd=00003D-vF/2, subfigure}
(note that for the proposed system configuration only pairwise entanglement
needs to be considered since coupling with the third qubit is so weak
that we can assume it doesn\textquoteright t exist), for the graphene
biased such that $v_{d}=\nicefrac{-v_{F}}{2}$, QB1 is entangled with
QB2, however, QB1 is not entangled with QB3. Now, if the graphene
is biased such that $v_{d}=\nicefrac{v_{F}}{2}$, as seen in Fig.
\ref{fig:entanglement control, vd=00003DvF/2, subfigure}, then QB1
is not entangled with QB2, however, QB1 is entangled with QB3.

\section{Conclusion}

We investigated DC current induced nonreciprocal graphene plasmon
polaritons as a candidate for entanglement mediation for enhancement
over vacuum. We used concurrence as a measure of entanglement. It
was shown that biasing the graphene sheet with a DC current induces
a nonreciprocal response with highly directed energy, which can be
used for entanglement enhancement and control. We have shown that
there was good entanglement enhancement over vacuum and that the proposed
configuration can be used to control which qubits are entangled by
changing the polarity of the DC bias.

\begin{backmatter} \bmsection{Funding} This work was supported
by the NATO Science for Peace and Security project\\
NATO.SPS.MYP.G5860.

\bmsection{Disclosures} The authors declare no conflicts of interest.

\bmsection{Data availability} Data underlying the results presented
in this paper are not publicly available at this time but may be obtained
from the authors upon reasonable request.

\end{backmatter}

\bibliography{Entanglement_mediated_by_DC_current_induced_NR_graphene_plasmonics}

\end{document}